\documentclass[preprint]{aastex}
\usepackage{gensymb}
\usepackage{amsmath}
\usepackage{mathtools}
\usepackage{graphicx}
\usepackage[section]{placeins}
\usepackage{hyperref}

\slugcomment{Draft of \today}

\shorttitle{The Quintuplet Proper Members}
\shortauthors{Hankins et al.}

\newcommand{\angstrom}{\mbox{\normalfont\AA}}

\begin{document}

\title{Infrared Observations of the Quintuplet Proper Members using SOFIA/FORCAST and Gemini/TReCS}

\author{M. J. Hankins$^1$, R. M. Lau$^{2,1}$, M. R. Morris$^3$, J. Sanchez-Bermudez$^{4,5}$, J. U. Pott$^4$, J. D. Adams$^{6,1}$, T. L. Herter$^1$}

\altaffiltext{1}{Astronomy Department, 202 Space Sciences Building, Cornell University, Ithaca, NY 14853-6801, USA}
\altaffiltext{2}{Jet Propulsion Laboratory, California Institute of Technology, 4800 Oak Grove Drive, Pasadena, CA, 91109-8099, USA}
\altaffiltext{3}{Department of Physics and Astronomy, University of California, Los Angeles, CA 90095-1547, USA}
\altaffiltext{4}{Max Planck Institut fur Astronomie, K\"{o}nigstuhl 17, 69117 Heidelberg, Germany}
\altaffiltext{5}{Instituto de Astrof\`{i}sica de Andaluc\`{i}a (CSIC), Glorieta de la Astronom\`{i}a S/N, 18008 Granada, Spain}
\altaffiltext{6}{Stratospheric Observatory for Infrared Astronomy, Universities Space Research Association, NASA/Armstrong Flight Research Center, 2825 East Avenue P, Palmdale, CA 93550, USA}

\begin{abstract}
Since their discovery, the Quintuplet proper members (QPMs) have been somewhat mysterious in nature. Originally dubbed the ``cocoon stars" due to their cool featureless spectra, high-resolution near-infrared imaging observations have shown that at least two of the objects exhibit ``pinwheel'' nebulae consistent with binary systems with a carbon-rich Wolf-Rayet star and O/B companion. In this paper, we present 19.7, 25.2, 31.5, and 37.1 $\mu$m observations of the QPMs (with an angular resolution of 3.2-3.8") taken with the Faint Object Infrared Camera for the SOFIA Telescope (FORCAST) in conjunction with high-resolution ($\sim$0.1-0.2") images at 8.8 and 11.7 $\mu$m from the Thermal-Region Camera Spectrograph (TReCS). DUSTY models of the thermal dust emission of two of the four detected QPMs, Q2 and Q3, are fitted by radial density profiles which are consistent with constant mass loss rates ($\rho_d \propto r^{-2}$). For the two remaining sources, Q1 and Q9, extended structures ($\sim 1"$) are detected around these objects in high-resolution imaging data. Based on the fitted dust masses, Q9 has an unusually large dust reservoir ($\mathrm{M_d}=1.3^{+0.8}_{-0.4}\times 10^{-3} \mathrm{M_{\odot}}$) compared to typical dusty Wolf-Rayet stars which suggests that it may have recently undergone an episode of enhanced mass loss.

\end{abstract}


\section{Introduction}

Five very luminous infrared sources reside in the young, massive Quintuplet cluster in the Galactic center (Okuda et al. 1990). These five objects, for which the cluster is named, are referred to as the Quintuplet proper members (QPMs). The QPMs are sources of great interest, given their enigmatic nature exhibiting high infrared luminosities  ($\sim10^{4}-10^5$ $\mathrm{L}_\odot$), cool spectral energy distributions ($400-800$ K), and previous claims of featureless near-infrared (IR) spectra (Moneti et al. 2001; Okuda et al. 1990; Nagata et al. 1990). Figer et al (1999) first proposed that the QPMs might be extremely dusty carbon-rich Wolf-Rayet stars (WCLd), but were unable to confirm this hypothesis without the identification of spectral features in the near-IR. From observations of the QPMs with the Infrared Space Observatory Camera (ISOCAM), Moneti et al. (2001) further explored this possibility, noting that the high IR luminosities ($\sim10^{4}-10^5$ $\mathrm{L}_\odot$) and small geometric dust covering fraction ($\sim$0.1) would be consistent with typical WC star luminosities; however, the lack of near-IR features in the J-band was still somewhat of a mystery. 

More recent multi-epoch, high-resolution near-IR images of two of the QPMs taken by the Keck 1 telescope have revealed ``pinwheel'' shaped plumes that are typically associated with colliding-wind binary systems (Tuthill et al. 2006). In the interpretation of Tuthill et al. (2006), the QPMs are late-type carbon-rich Wolf Rayet stars with an O/B star binary companion that form dust in the wind-wind collision front that is swept outward by the Wolf-Rayet winds (Tuthill et al. 1999). The classification of the QPMs as dust-enshrouded, evolved massive stars is consistent with the age of the Quintuplet cluster $\sim$3.3-3.6 Myrs (Liermann et al. 2012), which would be difficult to reconcile if the QPMs were young stellar objects (YSOs) as was originally suggested (Okuda et al. 1990). Although Tuthill el al. (2006) provide convincing evidence of the `pinwheel' nature of two of the QPMs: Q2 and Q3 (GCS 3-2 and GCS 4), the identification of the remaining members (Q1, Q4, and Q9 or alternatively, GCS 3-4, GCS 3-1, and GCS 3-3) was less certain. However, recent near-IR spectra taken by Geballe et al. (2014) clearly show broad emission line features associated with WC stars from Q1 and Q4 but not Q9. 

The two near-IR pinwheel stars, Q2 and Q3, appear to share somewhat similar properties. Blackbody fits to the SEDs of each object from Moneti et al. (2001) yield temperatures for Q2 and Q3 of 650 and 625 K. From the near-IR images, the systems appear to have, at least roughly, a similar geometry and are close to face-on (Tuthill et al. 2006). Q2 shows non-thermal radio emission and potentially weak x-ray emission (Lang et al. 2005; Law \& Yusef-Zadeh 2004) which is a telltale sign of a colliding wind system (Monnier et al. 2002). Q3 has no detected radio emission but exhibits stronger x-ray emission than Q2 (Law \& Yusef-Zadeh 2004). The remaining QPMs show no detectable radio or x-ray emission. Previous high-spatial-resolution images ($\sim 1''$) at 8.7 and 11.7 $\mu$m taken with the Palomar telescope show that the nebulae associated with Q9 and Q1 appear larger in physical extent than the other QPMs at low intensity levels and also display characteristically cooler spectra (400 and 500 K) compared with Q3 and Q2 (Moneti et al. 2001). Conversely, Q4 appears warmer (725 K) and more compact than Q2 and Q3.

In this paper, we present 19.7, 25.2, 31.5, and 37.1 $\mu$m observations of the QPMs taken by FORCAST aboard the Stratospheric Observatory for Infrared Astronomy (SOFIA). These observations provide important information on the mid-IR portion of the spectral energy distribution (SED) of the QPMs which can be modeled to provide estimates of dust production and mass loss rates of these objects. Additionally, we improve on previous high-resolution mid-IR imaging with 8.8 and 11.7 $\mu$m images from TReCS in full-pupil and Sparse Aperture Masking (SAM) mode to study the morphology of the dust reservoirs in these objects.  In a broader context, we aim to understand the effects of binarity in these systems as a study of massive evolved stars through their mass loss history. In a recent paper by Sana et al. (2012), it is suggested that $\sim$70\% of massive stars will exchange mass or merge with a binary companion over the course of their lifetime (see also Sana et al 2014). Thus, characterization of massive evolved binary systems is vital to understand the end states of massive stars. The objective of this analysis is to better classify these sources in terms of their thermal dust emission and offer possible explanations for their various similarities and differences.

\section{Observations and Data Reduction}

\subsection{SOFIA/FORCAST}

Observations of the QPMs were made using FORCAST (Herter et al. 2012) on the 2.5 m telescope aboard SOFIA. FORCAST is a $256 \times 256$ pixel dual-channel, wide-field mid-infrared camera sensitive from $5 - 40~\mu\mathrm{m}$ with a plate scale of $0.768''$ per pixel and field of view of $3.4'\,\times\,3.2'$.
The two channels consist of a short wavelength camera (SWC) operating at $5 - 25~\mu\mathrm{m}$ and a long wavelength
camera (LWC) operating at $28 - 40~\mu\mathrm{m}$. An internal dichroic beam-splitter enables simultaneous observation from both long and short wavelength cameras. A series of bandpass filters is used to image at selected wavelengths.

SOFIA/FORCAST observations of the QPMs were made on the Basic Science Flight 64 on June 8, 2011 (altitude $\sim$ 43,000 ft.) at 19.7, 31.4, and 37.1 $\mu\mathrm{m}$. Follow-up images of both the QPMs were taken on the OC1-B Flight 110 on July 1, 2013 (altitude $\sim$ 39,000 ft.) at 19.7, 25.2, 31.4, and 37.1 $\mu\mathrm{m}$. Measurements at 19.7 and 31.4 $\mu$m, as well as 25.2 and 31.4 $\mu$m for Flight 110,  were observed simultaneously using the dichroic beamsplitter, while the 37.1 $\mu\mathrm{m}$ observations were observed directly. Chopping and nodding were used to remove  the sky and telescope thermal backgrounds. An asymmetric chop pattern was used to place the source on the telescope axis, which eliminates optical aberrations (coma) on the source. The chop throw was $7'$ at a frequency of $\sim 4$ Hz. The off-source chop fields (regions of low mid-infrared Galactic emission) were selected from the Midcourse Space Experiment (MSX) $21~\mu$m image of the Galactic Center.

The source was dithered over the focal plane to allow removal of bad pixels and mitigate response variations.
The integration time at each dither position was $\sim 20$ sec. The quality of the images was consistent with
near-diffraction-limited imaging at $19.7 - 37.1~\mu$m; the full width at half maximum (FWHM) of the point spread function (PSF) was
$3.2''$ at $19.7~\mu$m, $3.4''$ at $25.2~\mu$m, $3.6''$ at $31.5~\mu$m,and $3.8''$ at 37.1 $\mu$m.

\begin{figure}[ht]
\centering
\includegraphics[width=70mm,scale=1.0]{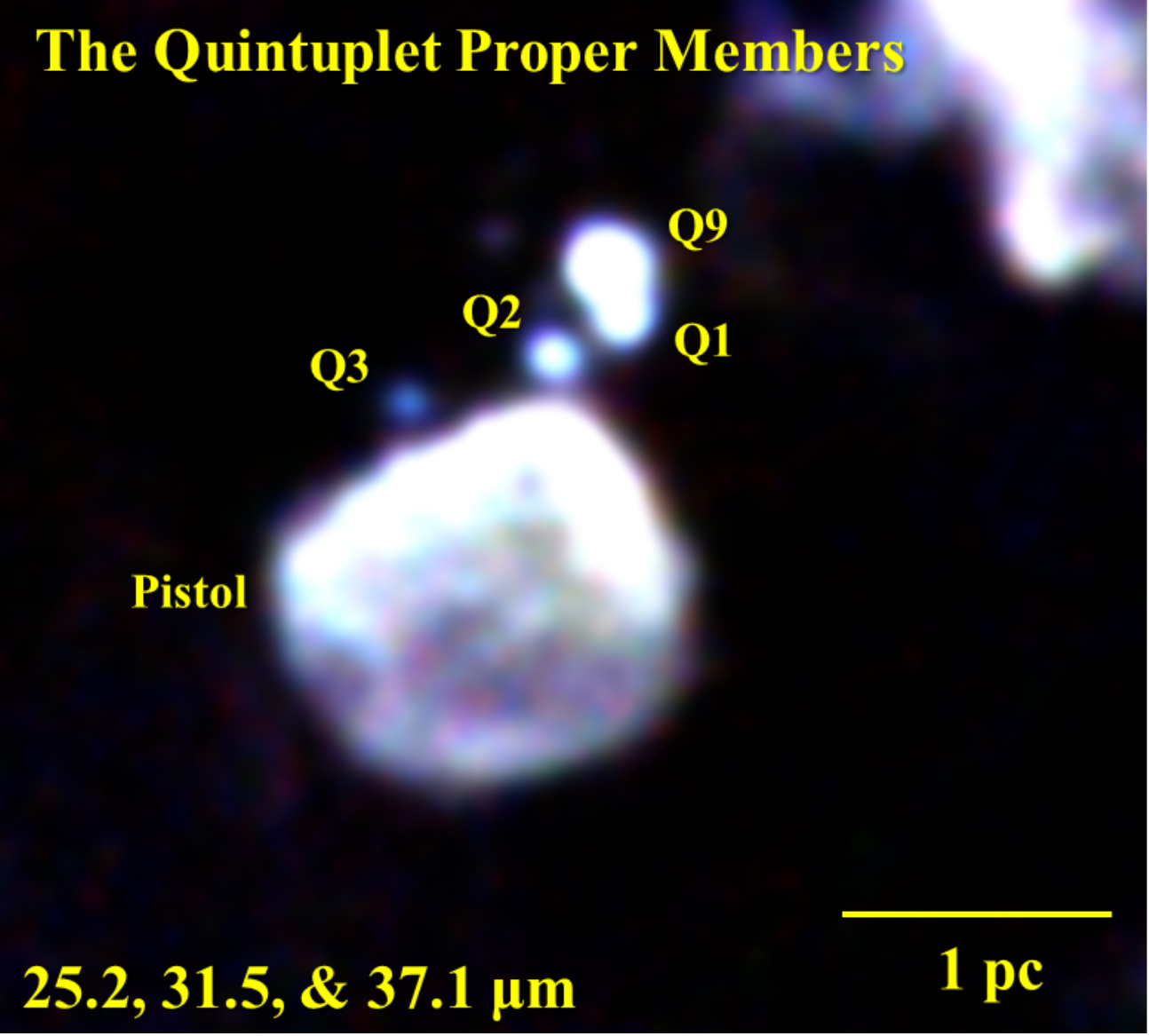}
\caption{ FORCAST 25.2 (blue), 31.5 (green), and 37.1 (red) $\mu$m false color image of the Quintuplet Proper Members. Q4 is only weakly detected at 19 $\mu$m.}
\label{fig:fig1}
\end{figure}

The acquired 240 s integrated images were reduced and combined at each wavelength according to the pipeline steps described
in Herter et al. (2013). Calibration was determined from the average response taken over the OC1-B Flight Series, adjusted to those of a flat spectrum ($\nu F_\nu = \rm{constant}$) source. The $3\sigma$ uncertainty in the calibration is estimated to be $\pm20\%$. The RMS noise per pixel was $26$ mJy at $19.7~\mu$m, $37$ mJy at $25.2~\mu$m, $17$ mJy at $31.5~\mu$m, and $26$ mJy at 37.1 $\mu$m. The FORCAST images were deconvolved using the Richardson-Lucy algorithm (Richardson 1972, Lucy 1974) with the calibrator $\mathrm{\beta}$ Umi used to estimate the PSF. The images were then convolved back to a uniform resolution of $3.4''$. Figure \ref{fig:fig1} displays the combined data from three bands in a false color image using the 25.2, 31.5, and 37.1 $\mu$m filters. Only four of the five QPMs were detected in observations of all four bands.\footnote{Q4 was weakly detected at 19 $\mu$m, and may be present in other bands but lacks sufficient SNR to cleanly extract.}

\subsection{Gemini/TReCS}

\begin{figure}[hb]
\centering
\includegraphics[width=90 mm]{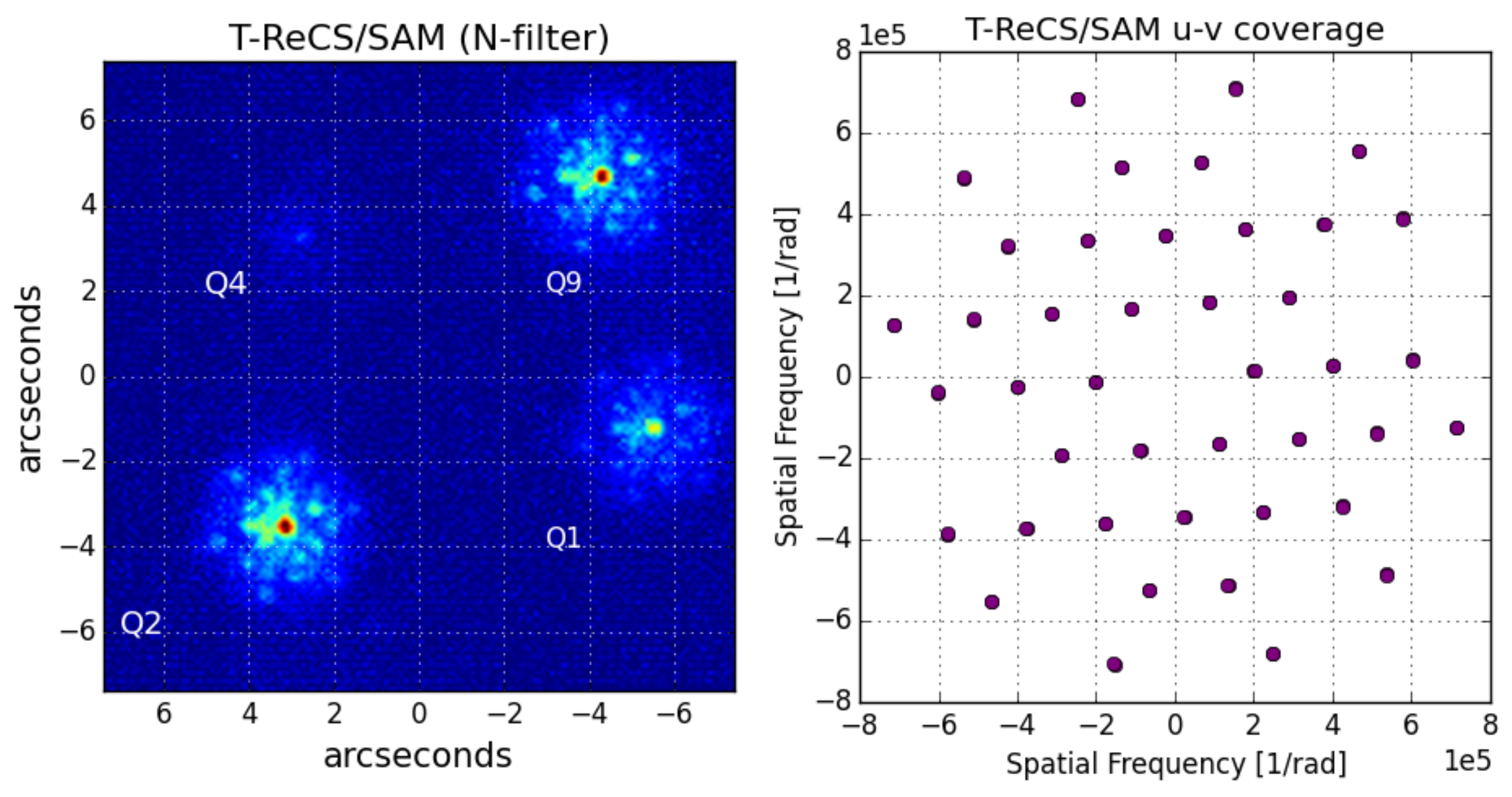}
  \caption{The \textit{left} panel displays one example of the
    Quintuplet stars Q4, Q2, Q9 and Q1. On the \textit{right} panel the
  \textit{u-v} coverage of the interferometric observations is
  displayed. It has 42 points which correspond to the 21 baselines
  (and their conjugates)
  formed with the \textit{7holes} mask.}
\label{fig:uv_coverage}
\end{figure}

Observations of the QPMs taken with the Thermal-Region Camera
Spectrograph (T-ReCS) on the night of September 27, 2007 were
available on the Gemini Science Archive. The data consist of
 several  data cubes taken with standard full-pupil imaging, and with
 the Sparse Aperture Masking (SAM) mode of TReCS. Table \ref{tab:obs}
displays different data sets used in this work. For the standard data
reduction, custom IDL routines were used. First, each one of the frames
in the different data sets was bad-pixel corrected. 
Second, since the cosmetic flaws of the detector presented a series of vertical stripes with a
width of $\sim$30 pixels parallel to the vertical axis, the striping 
artifacts in the data were corrected by replacing these values with an average of the pixels that correspond to the first and last 
stripes observed in the detector field of view (FOV). Finally, an average over the frames of 
the second nodding position in each data cube was subtracted from each frame of the first nodding 
position to minimize the effects of the sky background. Figure \ref{fig:uv_coverage} displays an average interferogram 
for one of the data cubes observed. The components of the Quintuplet system 
that are captured in the instrument field of view are labeled.\footnote{Q3 was outside of the field of view for these observations} 
Additionally, this figure also exhibits the \textit{u-v}
coverage of the interferometric observations.

\begin{figure}[htp]
\centering
\includegraphics[width=12 cm]{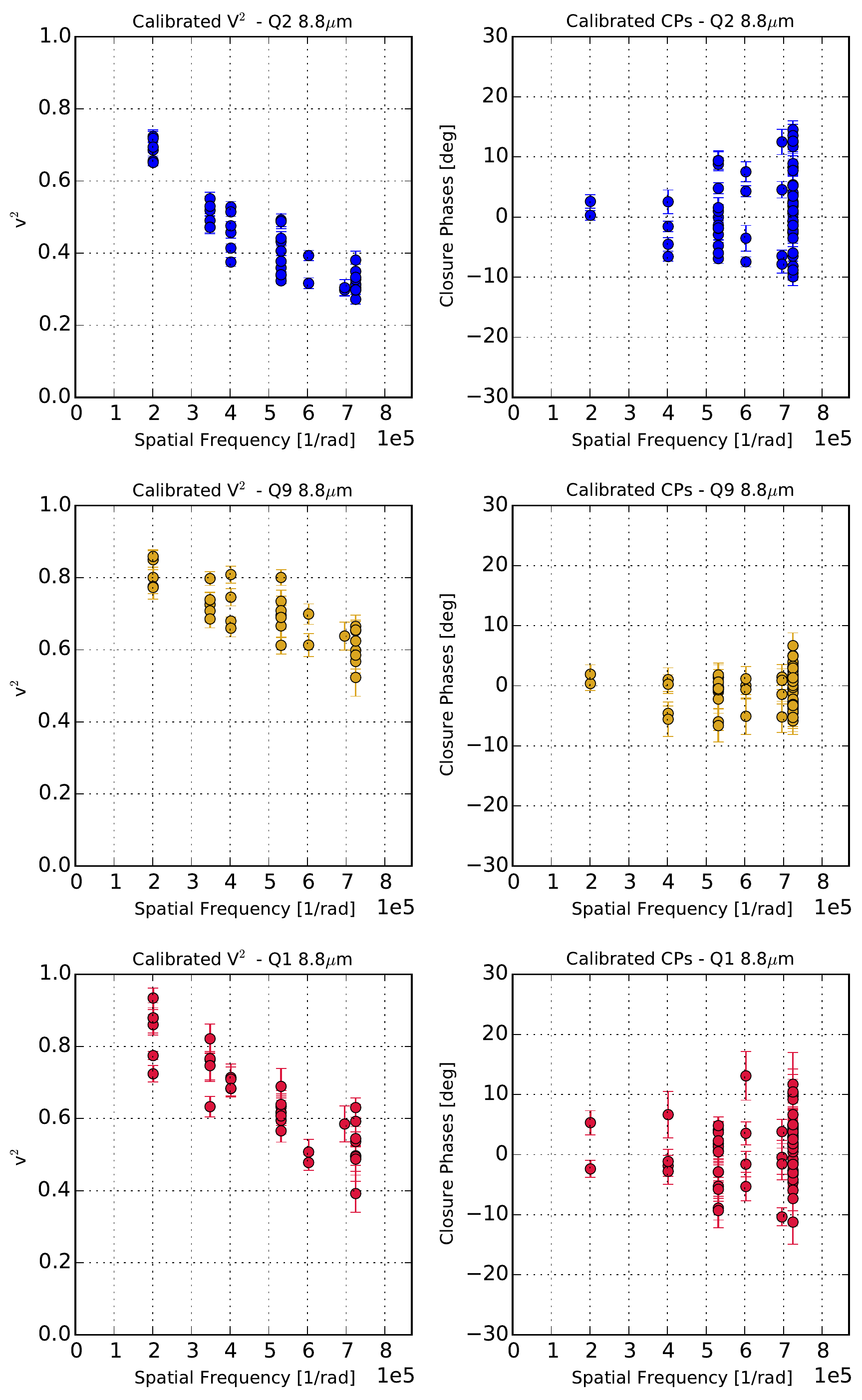}
  \caption{The \textit{left} column displays the calibrated squared
    visibilities obtained with the T-ReCS/SAM observations at 8.8 $\mu$m of the
    Quintuplet stars: Q2, Q9 and Q1. On the \textit{right} column the
    calibrated closure phases are displayed. The data of the different stars are
    identified with different colors}
\label{fig:io_n88}
\end{figure}

\begin{figure}[htp]
\centering

\includegraphics[width=12 cm]{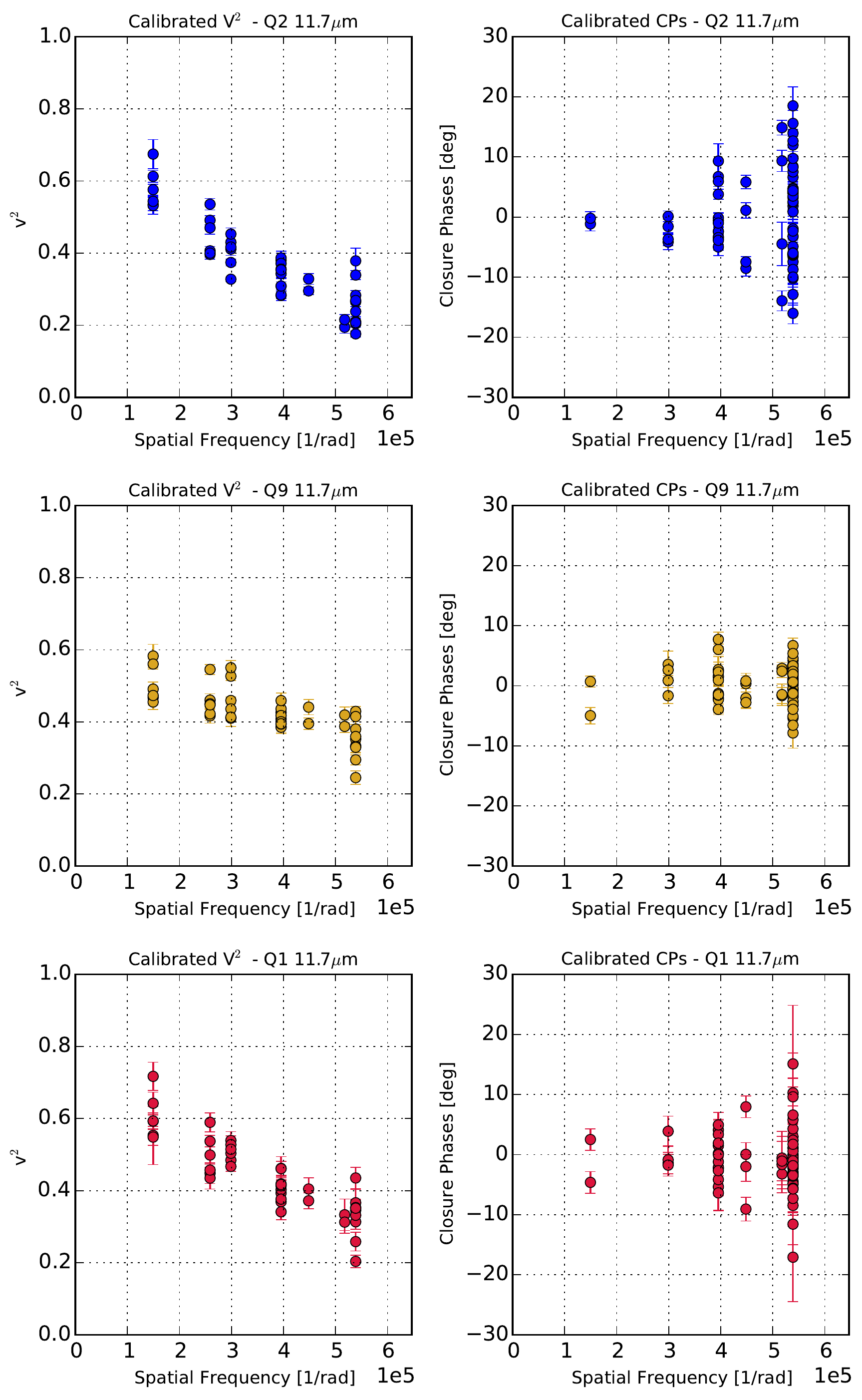}
  \caption{The \textit{left} column displays the calibrated squared
    visibilities obtained with the T-ReCS/SAM observations at 11.7 $\mu$m of the
    Quintuplet stars: Q2, Q9 and Q1. On the \textit{right} column the
    calibrated closure phases are displayed. The data of the different stars are
    identified with different colors}
\label{fig:io_n117}
\end{figure}

Aperture Masking is a technique that transforms a telescope into a
Fizeau interferometer by placing a mask with some holes at the pupil
plane. With this observational model, what is recorded at the camera detector is the transmission function of the
interferometer convolved with the real brightness distribution of the
observed targets. This pattern is known as an
\textit{interferogram}, and the observables (squared visibilities and
closure phases) are obtained via its Fourier analysis. To obtain the
interferometric observables, the SAM data were analyzed with the IDL data reduction software
developed at the University of Sydney (Tuthill et al., 2000,
2006). For both filters, the visibilities and closure phases were
calibrated using an average of the visibilities obtained with the two
observed SAM calibrators. Figures \ref{fig:io_n88} and \ref{fig:io_n117} display the calibrated
interferometric observables for targets Q2, Q9 and Q1 at 8.8 and 11.7
$\mu$m, respectively.

 As observed, all the targets are at least partially resolved
 in both filters. Closure phases vary from around -20 to 20
 degrees, which suggests that the morphology of the targets is quite
 symmetric. Nevertheless, small variations in the closure phases
 might be a consequence of the partially resolved structure since the 
 angular resolution of the SAM data is larger than the angular size of the targets.

\begin{figure}[htp]
\centering
\begin{minipage}[c]{0.47\linewidth}
    \includegraphics[width=6 cm]{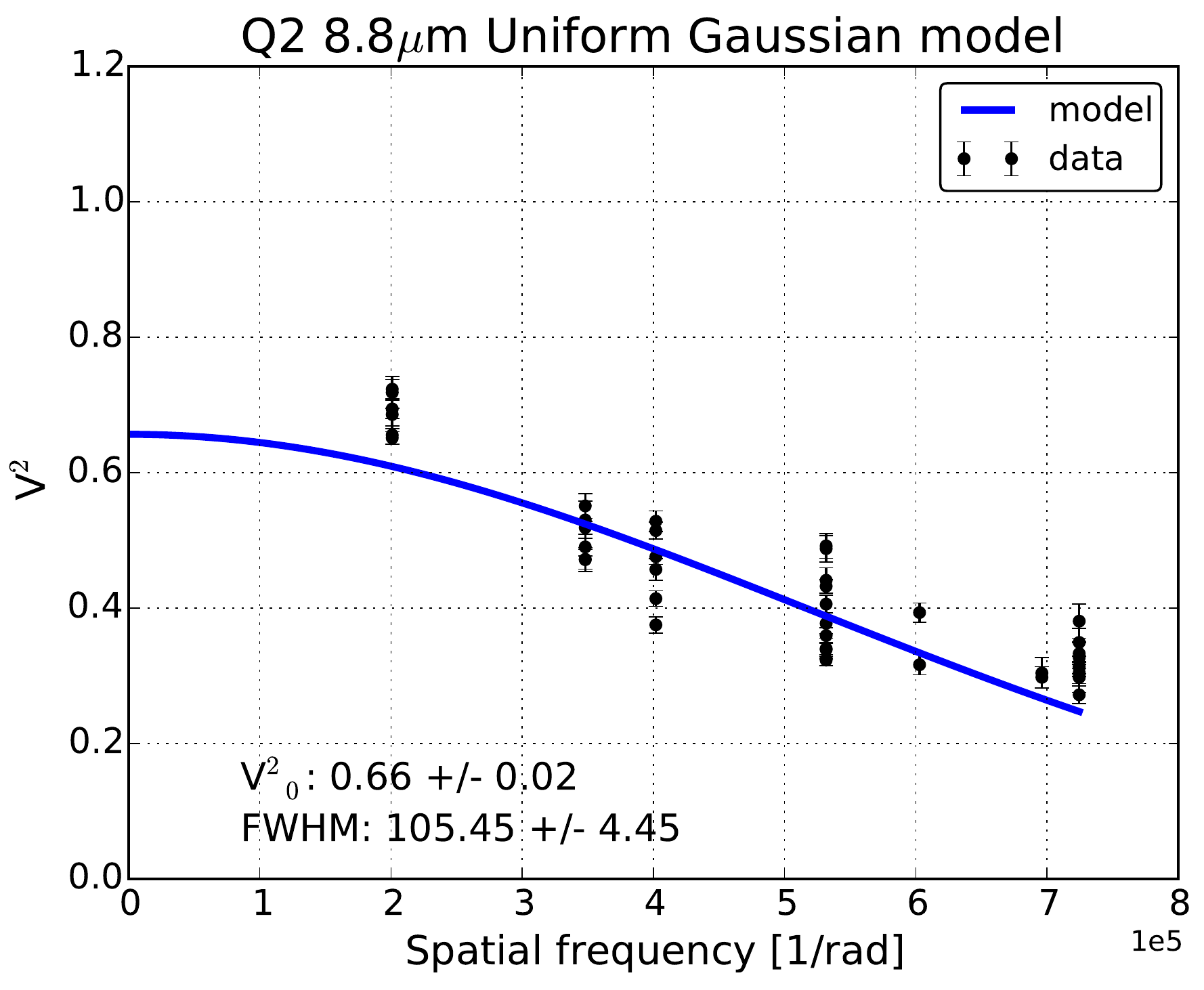}
\end{minipage}
\begin{minipage}[c]{0.47\linewidth}
    \includegraphics[width=6 cm]{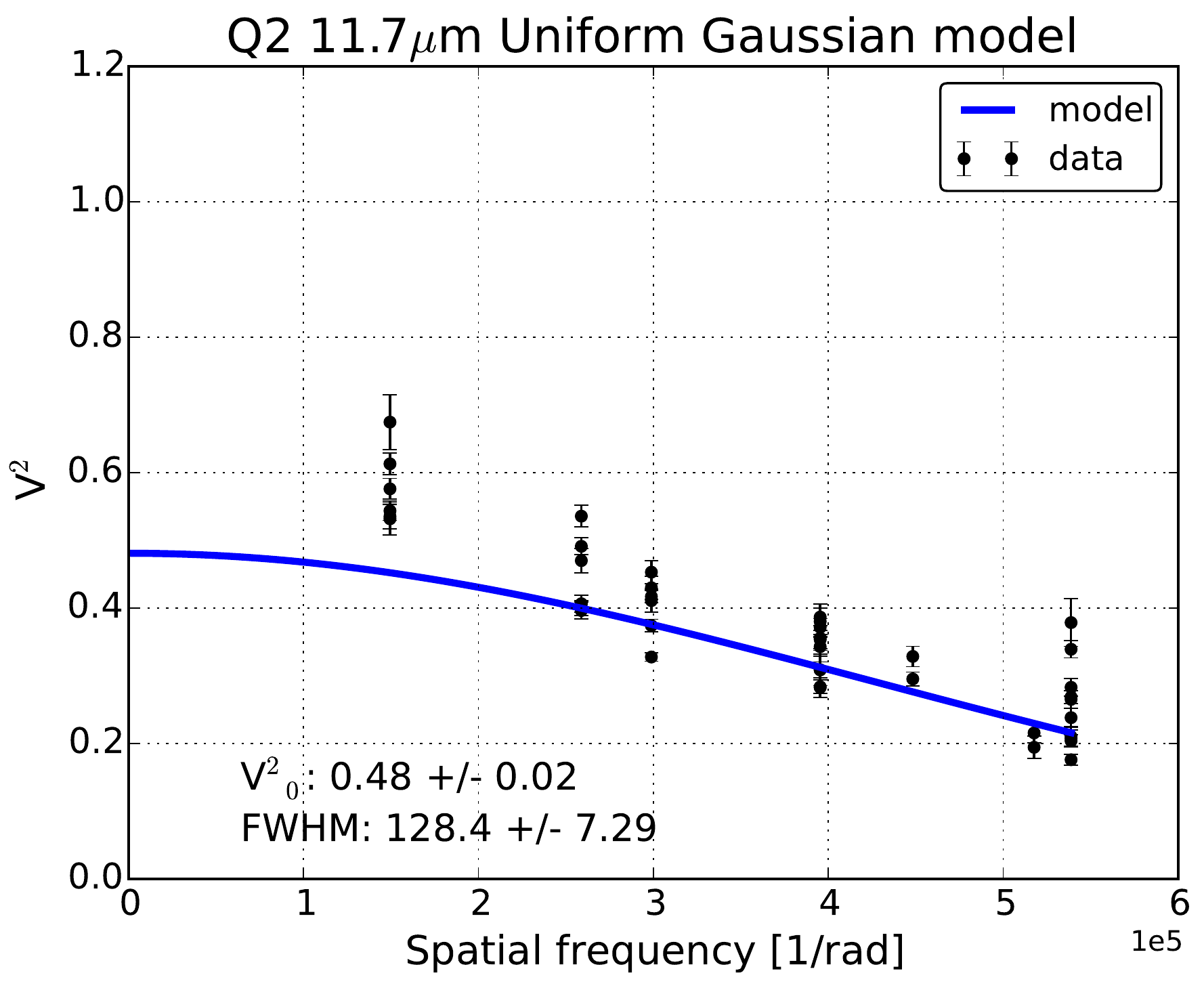}
\end{minipage}
\begin{minipage}[c]{0.47\linewidth}
    \includegraphics[width=6 cm]{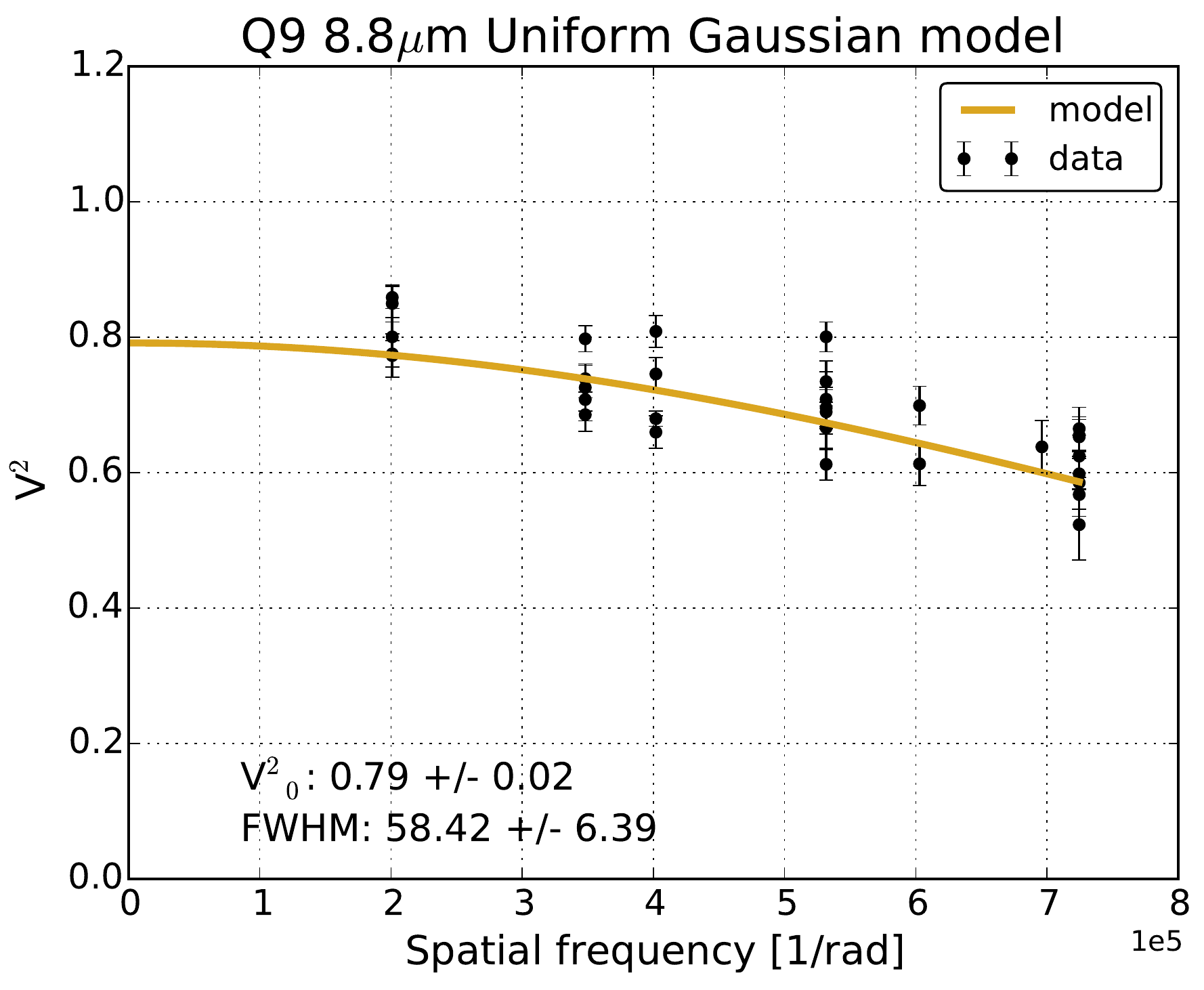}
\end{minipage}
\begin{minipage}[c]{0.47\linewidth}
    \includegraphics[width=6 cm]{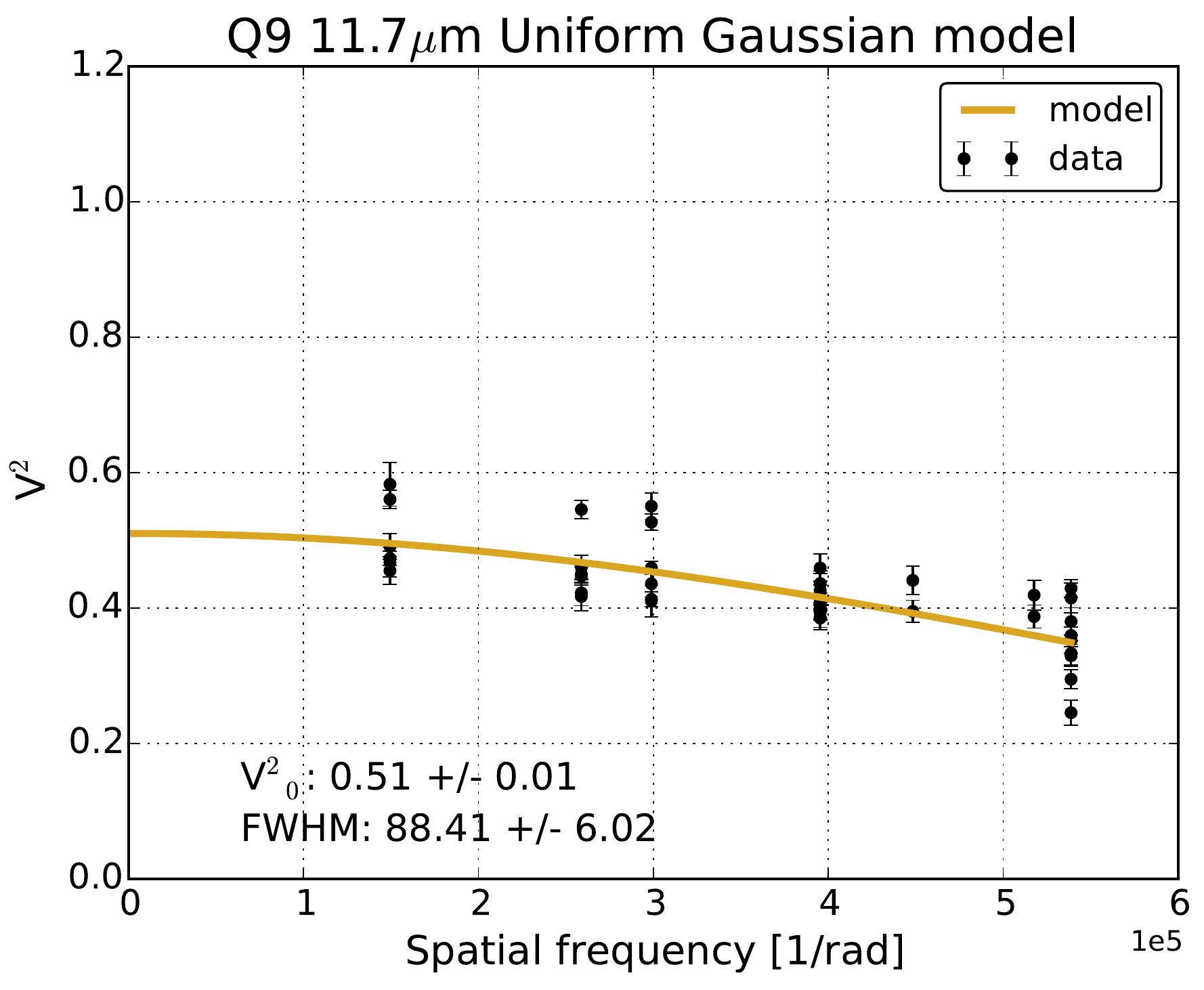}
\end{minipage}
\begin{minipage}[c]{0.47\linewidth}
    \includegraphics[width=6 cm]{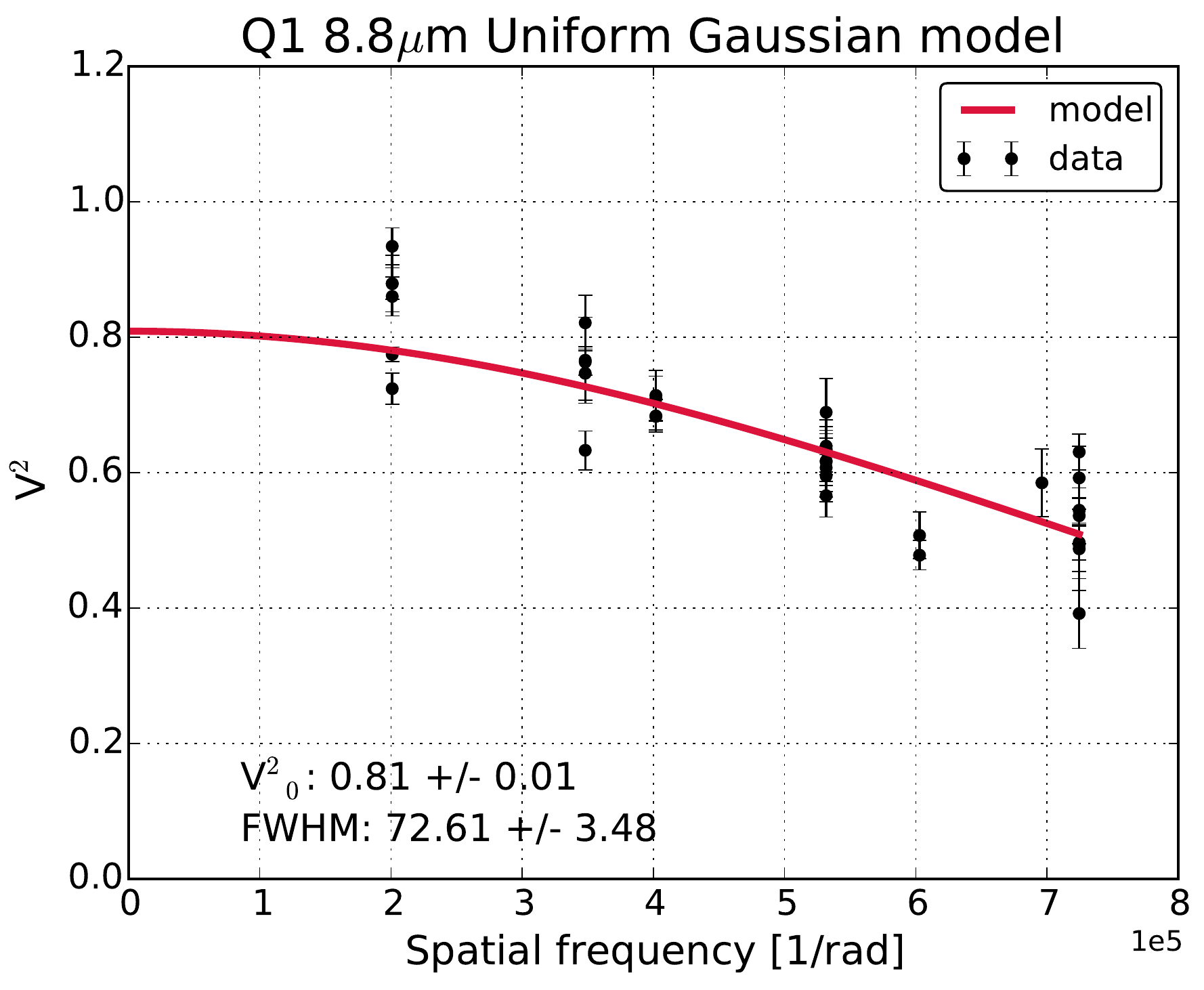}
\end{minipage}
\begin{minipage}[c]{0.47\linewidth}
    \includegraphics[width=6 cm]{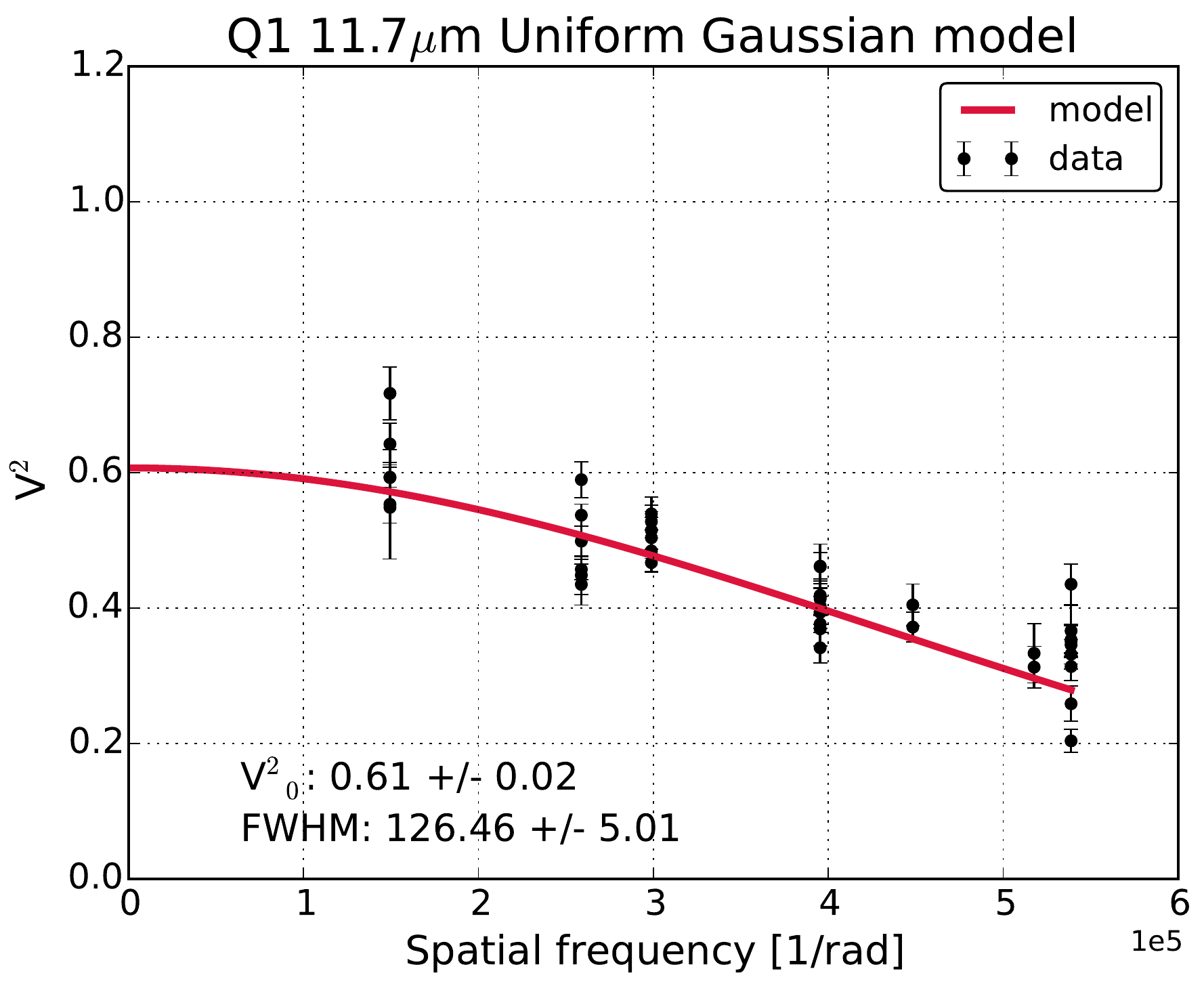}
\end{minipage}
  \caption{Best-fit Gaussian model to the
    $V^2$. The data are plotted in black circles and the model in
    colors, depending on the source. The best-fit values are also
    labeled. The parameter V$^2_0$ is equivalent to the value of the
    squared visibility at the zero-baseline.}
\label{fig:model_fit}
\end{figure}

A simple geometrical model of a Gaussian was fitted to the calibrated visibilities (V$^2$) to estimate the angular size of the target. Table \ref{tab:best-fit} displays the parameters of the best-fit models to the visibilities and Figure \ref{fig:model_fit} shows the best-fit models overplotted on the observed data. As it is observed in the best-fit values of the models, the value of
V$^2$ at zero-baseline is lower than unity. This fact suggests the presence
of an over-resolved flux for all the observed targets and filters, which
implies that the sources are more extended than the angular size that
could be measured with the shortest baseline of the interferometric configuration employed.

\begin{figure}[htp]
\centering
\begin{minipage}[c]{0.47\linewidth}
    \includegraphics[width=6 cm]{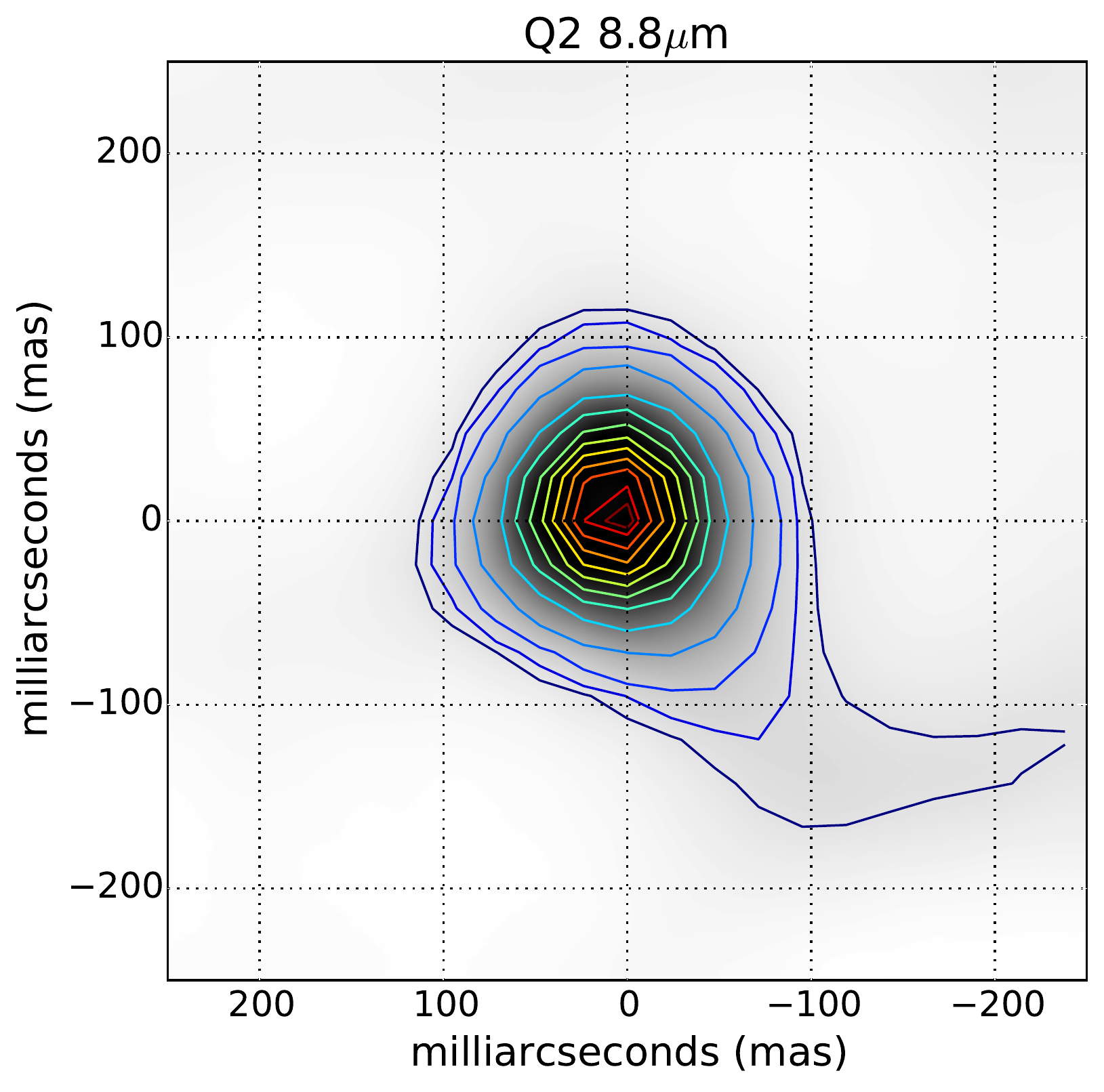}
\end{minipage}
\begin{minipage}[c]{0.47\linewidth}
    \includegraphics[width=6 cm]{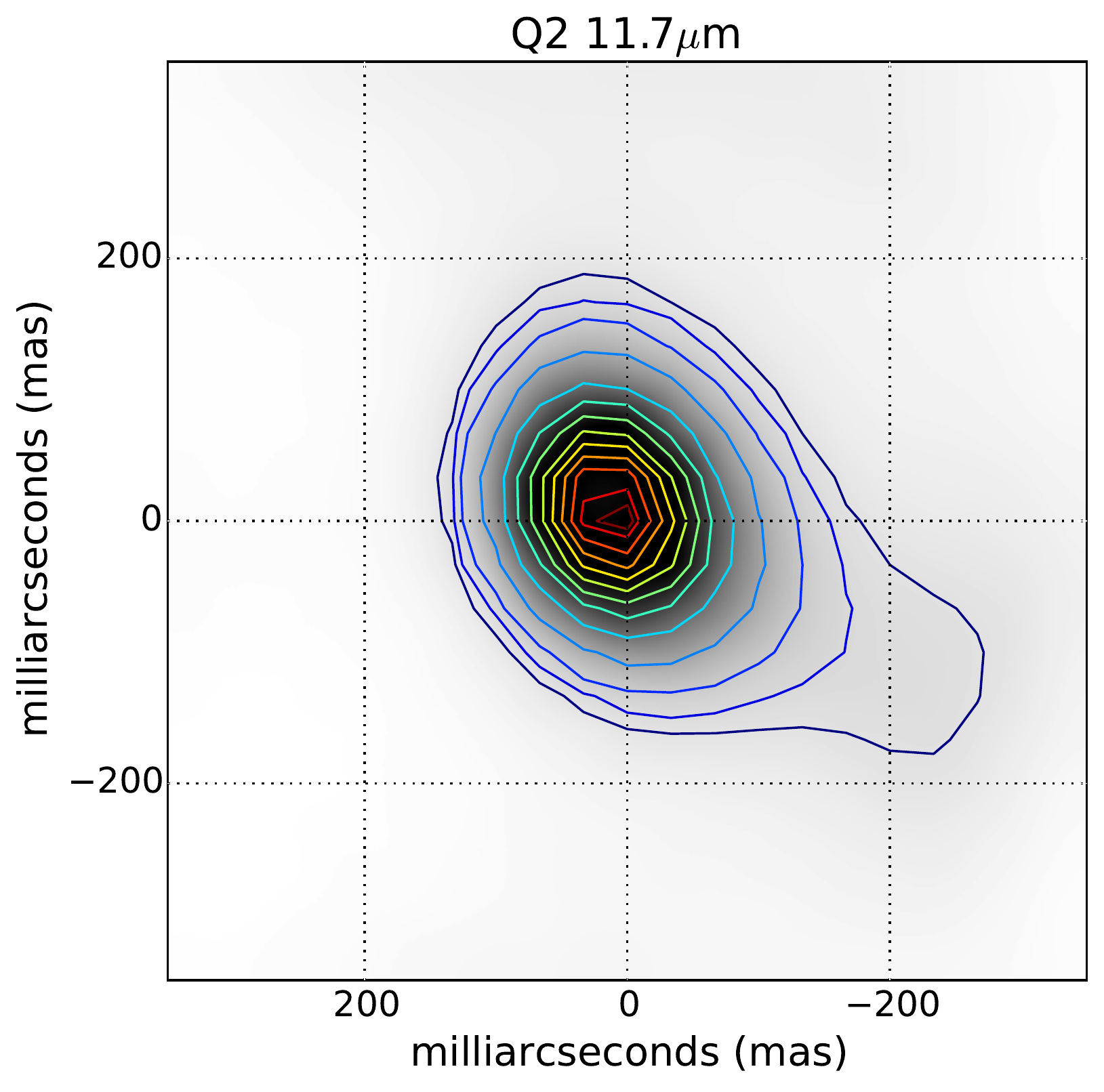}
\end{minipage}
\begin{minipage}[c]{0.47\linewidth}
    \includegraphics[width=6 cm]{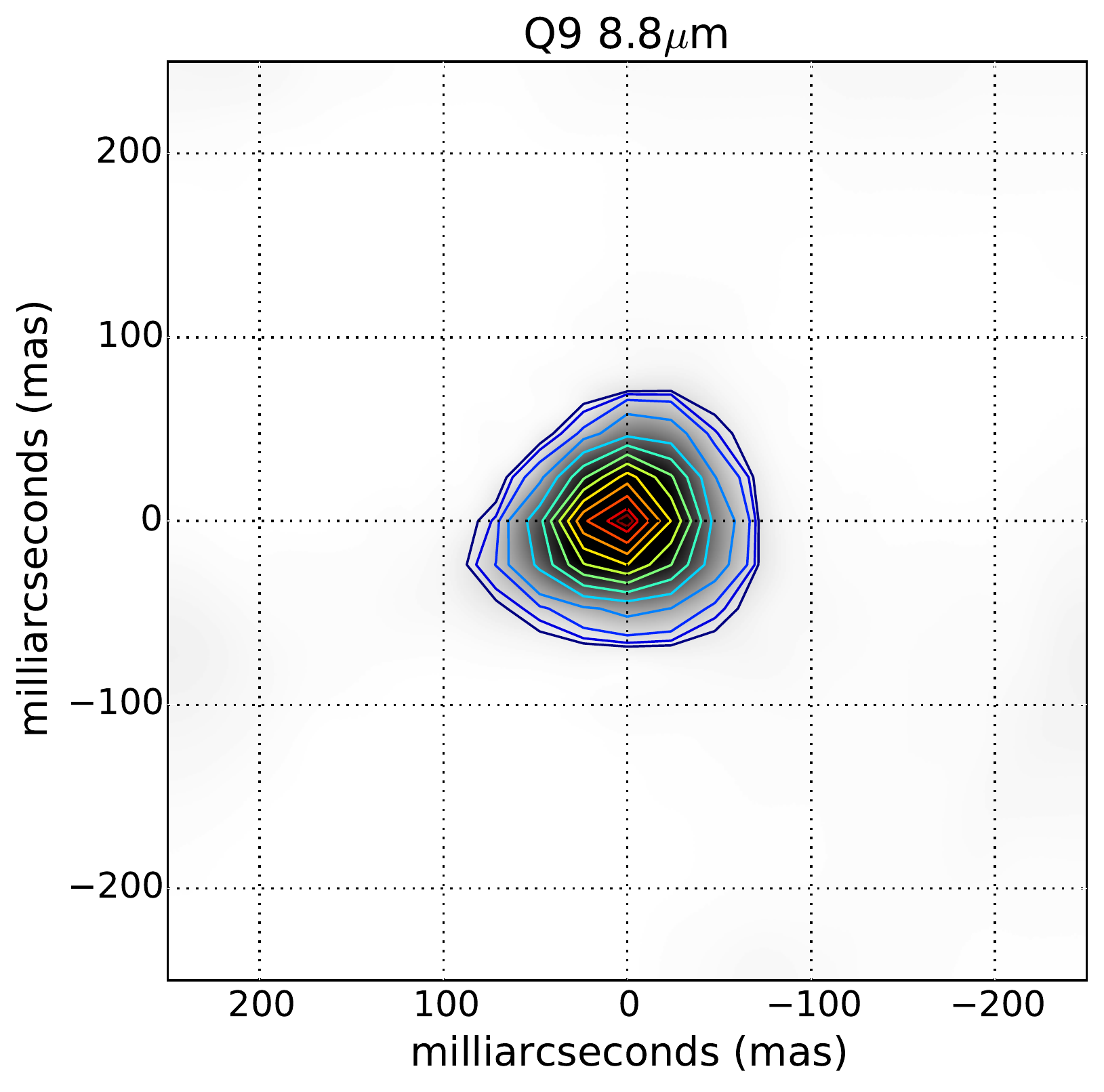}
\end{minipage}
\begin{minipage}[c]{0.47\linewidth}
    \includegraphics[width=6 cm]{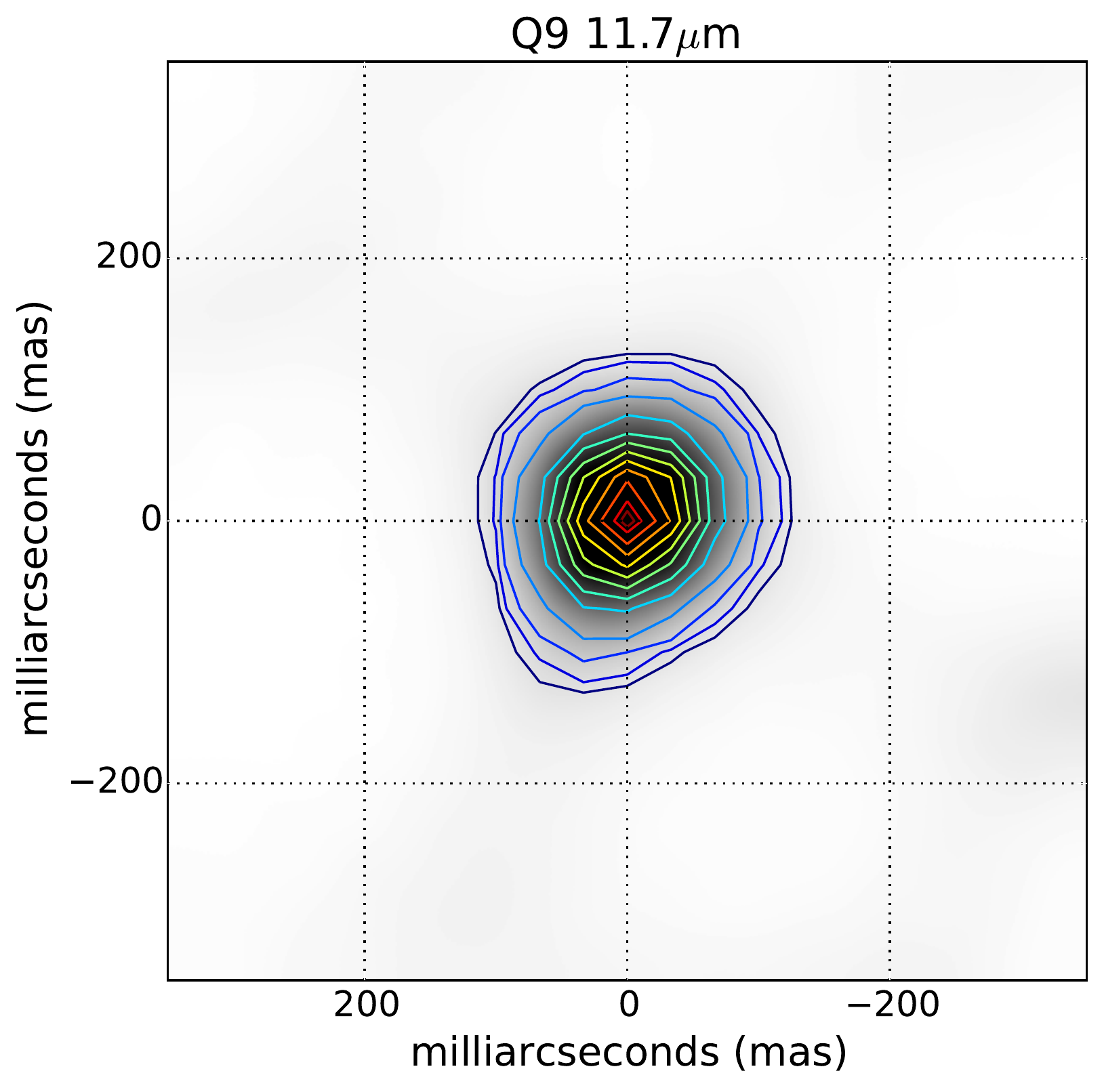}
\end{minipage}
\begin{minipage}[c]{0.47\linewidth}
    \includegraphics[width=6 cm]{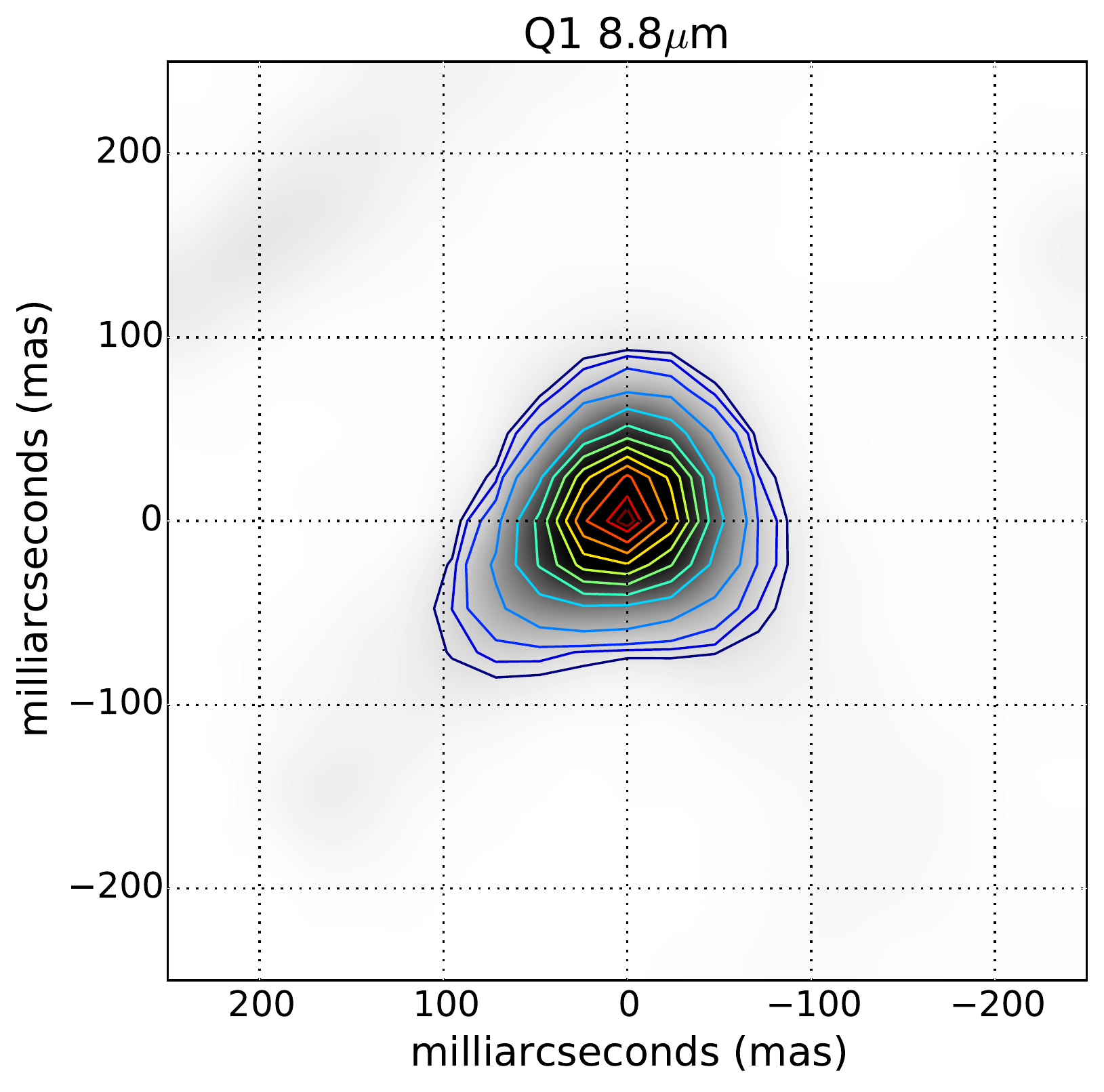}
\end{minipage}
\begin{minipage}[c]{0.47\linewidth}
    \includegraphics[width=6 cm]{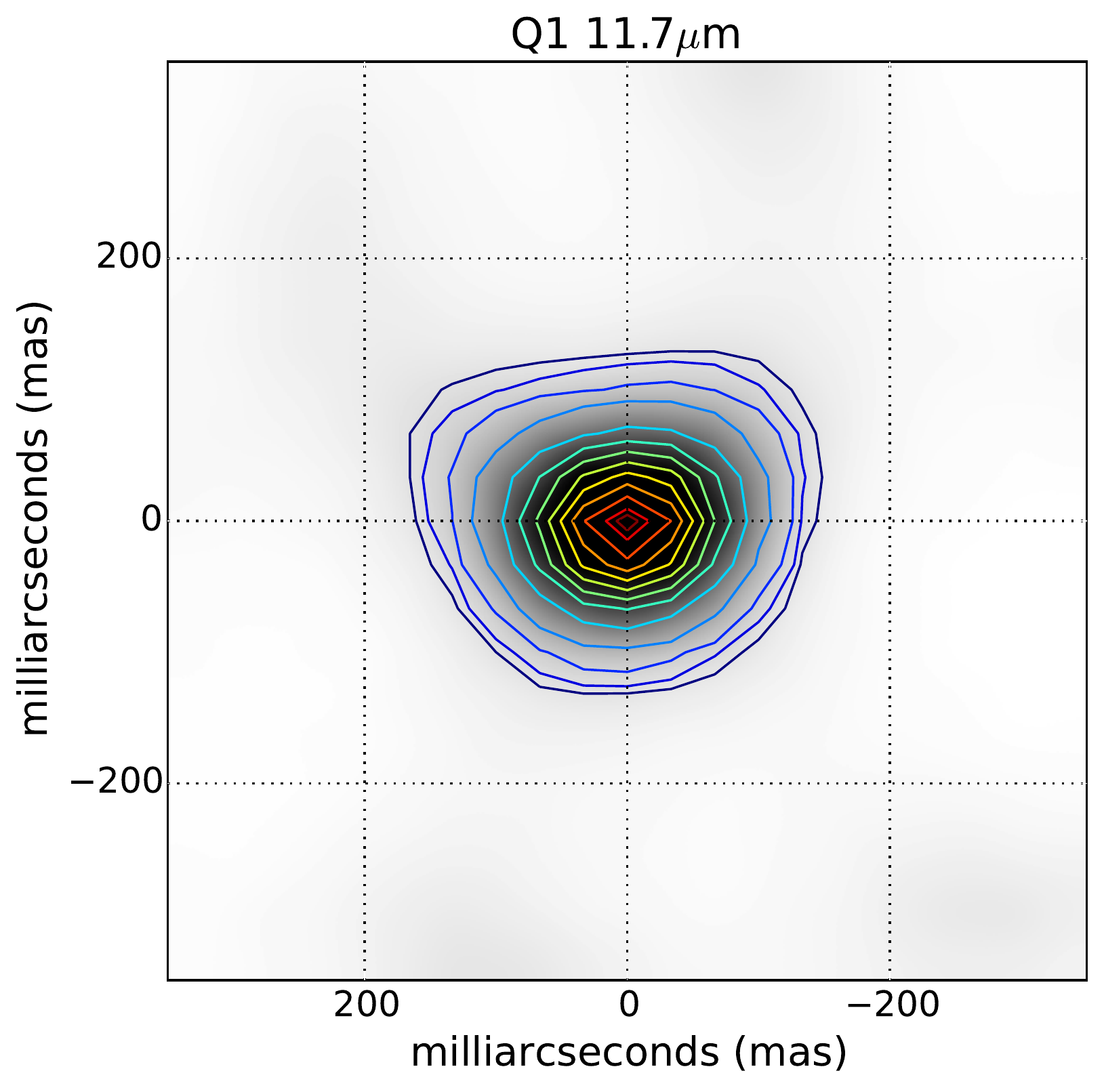}
\end{minipage}
  \caption{Reconstructed BSMEM images of the
    Q2, Q9 and Q1 sources. The  images display contour levels of 90, 80, 70, 60, 50, 40, 30, 20, 10, 5, 3, and 2 $\%$ of the peak flux}
\label{fig:bsmem_im}
\end{figure}

Image reconstruction of the targets was performed with BSMEM
(Buscher 1994). This software uses a maximum entropy algorithm to fit a model of the source's brightness distribution
directly to the closure phases and squared visibilities. The resolution of the reconstructed image depends on the quality (uncertainties) of the data and on the angular resolution obtained with the longest
baseline of the interferometric array (see Monnier 2003, 2012). In this case, the sources exhibit compact structures of the order of the nominal angular resolution of the interferometer ($\theta=\lambda/2\mathrm{D})$. Table
\ref{tab:bsmem_par} shows the parameters used for the image reconstruction process of the different targets, assuming an initial Lorentzian distribution with a FHWM=10 mas and a Gull-Skilling entropy regularizer. Figure \ref{fig:bsmem_im} displays the images obtained as a result of the reconstruction process.

To search for larger extended structures in the full-pupil data, StarFinder (Diolaiti et al. 2000) was used to subtract the central sources based on a model PSF. The compact unresolved source Q4 was used to estimate the PSF of the data for the subtraction process. Standard flux calibrators were not observed  together with the analysed data. Therefore, to flux calibrate the images, the observed fluxes were cross-calibrated with the flux values from previous Palomar mid-IR imaging (Moneti et al. 2001), which are consistent to $\sim$20\%.

\subsection{Data from Literature and Near-IR Variability}

Observations from a variety of sources in the literature were incorporated in this analysis to expand the wavelength coverage of the SEDs of the QPMs. J, H, and K band measurements of the QPMs were taken from the 2MASS point source catalog (Cutri et al. 2003; Skrutskie et al. 2006) to cover the near-IR portion of the spectra. Mid-IR fluxes at 3.6, 4.5, 5.8 and 8 $\mu$m of the QPMs were taken from the Spitzer/IRAC Survey of the Galactic Center Point Source Catalog (Ram\`{i}rez et al. 2008). Finally, because several dusty WC stars are known to be highly variable (e.g. WR48a (Williams et al 2012)), it was necessary to check the QPMs for variability. A near-IR variability study by Matsunaga et al. (2009) which included the Quintuplet region found that Q2 varies quite substantially over the course of a few years ($\sim$1 mag. in J band), and Q9 may also show some degree of variability but is less clear from the available data. Due to this fact, the measured fluxes at the time of the 2MASS (MJD 51824) and IRAC (MJD 53459) observations are too far apart to be consistent for highly variable sources. To construct a more consistent SED for Q2, near-IR photometry from Matsunaga et al. (2009) observed on MJD 53537\footnote{This is the closest available observation to the IRAC observations} was used for this object. Finally, because of the potential for low-level near-IR variability, fairly conservative 20\% uncertainties were adopted for the near-IR fluxes for all of the QPMs.

\section{Model Description and Parameters}

\subsection{Interstellar Extinction}

Large column densities of dust and gas lead to substantial extinction along lines of sight towards the Galactic center ($A_V\sim30$) (Cardelli et al. 1989). In this work, we assume a distance to the Quintuplet cluster consistent with the Galactic center- 8 kpc (Reid 1993). Additionally, we adopt the extinction law of Fritz et al. (2011) who derived an interstellar extinction curve from hydrogen recombination line observations of Sgr A West -- the HII region in the inner 3 pc of the Galactic center -- at 1 - 19 $\mu$m made by the Short Wave Spectrometer (SWS) on the Infrared Space Observatory (ISO) and the Spectrograph for Integral Field Observations in the Near Infrared (SINFONI) on the Very Large Telescope (VLT). In this analysis, the extinction correction was scaled to $\mathrm{A_K}=1.8$ which was determined by using the depth of the silicate feature in the QPMs. Since the mid-IR emitting portions of the circumstellar outflows of dusty WC stars are thought to be composed entirely of amorphous carbon dust, any silicate emission should be purely line of sight and not intrinsic to the objects themselves. The adopted extinction correction is slightly lower than the Fritz et al. (2011) value for the Galactic center ($\mathrm{A_K}=2.4$); however, it is well established that the sight line to the Quintuplet cluster suffers less extinction than other positions in the Galactic center (Hu{\ss}mann et al. 2012).

We studied the use of the extinction law from Chiar \& Tielens (2006), who derive an extinction law using ISO measurements of Q4 from 2.4-40 $\mu$m. In these observations, the large aperture size used (14" $\times$ 20" and 14" $\times$ 27") precludes the ability to resolve the emission from individual QPMs in the GCS 3 complex.\footnote{See figure 1 from Moneti et al. 2001} Most noticeably, the Chiar \& Tielens extinction law differs in the detailed shape of the 9.7 $\mu$m silicate feature from that of Fritz et al. (2011); however, the spectrum of Q4 used to derive it is potentially misleading, especially at longer wavelengths where the cooler emission of Q9 likely dominates over Q4. This may explain why their extinction law follows approximately a $\lambda^{-1}$ law rather than the generally assumed $\lambda^{-2}$ power law at wavelengths longer than 24 $\mu$m (Draine \& Lee 1984).

\subsection{DUSTY Models}

The radiative transfer code DUSTY (Nenkova et al. 2000) was used to model the detailed SEDs of the QPMs. To find the best-fit parameters, a grid of the various parameter values was constructed and the reduced $\chi^2$ statistic was determined for each model. The model fitting focused on adjusting three main parameters: the inner boundary temperature of the dust shell, $\mathrm{T_0}$, the radial size of the nebula, $R$, and the central source luminosity, L$_{\star}$. It was expected that the dust in the nebula likely follows a radial density power law of $r^{-2}$, consistent with a constant mass loss rate. However, the radial density profile was allowed to vary in two of the objects which were not well fit by the $r^{-2}$ dust density profile.

For grain properties, we assume dust parameters for the QPMs consistent with nebulae around other dusty WC stars. The dust is composed entirely of amorphous carbon, and three grain size populations were tested- small (0.01 $\mu$m), medium (0.1 $\mu$m), and large (1 $\mu$m) grains. The small grain sample is consistent with the calculations of Zubko (1998), who derived a characteristic grain size (100-200 $\angstrom$) for dust grains in Wolf-Rayet winds based on the balance between grain growth and destruction by sputtering in the winds. The large grain sample is consistent with observational work by Chiar \& Tielens (2001) and Marchenko et al. (2002) who suggest that dust grains in dusty WC stars could be as large as 0.5 to 1 $\mu$m.

There are several additional parameters that are needed for the DUSTY models. First, the input radiation field for the DUSTY model is modeled as a blackbody with T$_{\star}$=40,000 K. The value of T$_{\star}$ is fairly typical for WC9 subtypes (Sander et al. 2012) which make up the majority of the population of dusty WC stars (van der Hucht 2001); however, $\sim$30\% of dusty WC stars are earlier subtypes with much higher effective stellar temperatures. The observed continuum of the star (e.g. at J band) constrains the stellar luminosity as $F_{\nu} \sim L/T^3$ (in the Rayleigh-Jeans limit), so ultimately the uncertainty in T$_{\star}$ limits our ability to constrain L$_{\star}$ for these sources which varies by a factor of $\sim$2 between subtypes of late-type WC stars (Crowther 2007).

DUSTY uses a spherically symmetric geometry, which could be problematic for Q2 and Q3, which have known pinwheel-like disk structures in their inner regions. Additionally, the detailed geometries for Q1 and Q9 are unknown. However, as shown in Figure \ref{fig:tauPlot}, the emission of the models in the mid-IR is optically thin. This implies that the main concern is properly modeling the temperature of the dust as a function of distance from the star. For example, if the dust lies in a disk, the optical depth proceeding from the star outward is significantly larger than that of a spherical distribution for a fixed dust mass versus radius (the increase depends on the opening angle). Self-consistent models of colliding wind (pinwheel) disks give a temperature profile similar to that of the DUSTY model at distances outside of the first few spiral turns (Harries et al. 2004). Using this temperature profile for Q2 and Q3 has no effect on the analysis. Thus for consistency we have chosen to utilize DUSTY models for all sources.

\section{Results and Analysis}

\subsection{Best-Fit Model Parameters}

The best-fit DUSTY models for the QPMs are plotted in Figure \ref{fig:Qmodels} and the model parameters for each of the QPMs are listed in Table \ref{tab:SpiralBestFit}. The best-fit model parameters for the near-IR pinwheel QPMs (Q2 and Q3) are fairly similar. Both are consistent with an $r^{-2}$ dust density profile and have comparable inner boundary temperatures and sizes. The best-fit source luminosity of Q3 is $\mathrm{Log(L_{\star}/L_{\odot})}=5.5$; however, this quantity is not as well determined for Q2. The shape of the near-IR SED of Q2 is difficult to fit with the models, especially at H band (the J band flux is dominated by stellar photospheric emission). Because Q2 varies greatly in its near-IR magnitude, we suggest that it is likely a variable dust former. This hypothesis could also explain why Tuthill et al. (2006) were only able to observe the pinwheel around Q2 in one epoch. Additionally, this may create deviations in the amount of hot dust in close to the central system relative to our models which could explain poor fit to this object in the near-IR; however, the cooler reservoir appears to be consistent with a constant mass loss rate over the timescale of the observable dust component in the mid-IR. Assuming the J band flux is representative of the photospheric emission of the central source and a stellar temperature of 40,000 K, the apparent stellar contribution needed to match the SED of Q2 implies a source luminosity of $\mathrm{Log(L_{\star}/L_{\odot})\sim5.7}$.

The remaining QPMs (Q1 and Q9) have stronger emission relative to Q2 and Q3 beyond 20 $\mu$m, and are not well fit by a $r^{-2}$ dust density profile. In the case of these two objects, the density profile was varied which allowed for greater amounts of cool dust to fill in this portion of the SED. Q1 is fairly well fit by a $r^{-\alpha}$ density profile with $\alpha=1.\pm0.5$. Q9 shows evidence for a significantly greater mass loss rate in the past, it is best fit by a density profile with $r^{-\alpha}$ with $\alpha=-1.5\pm0.5$. For Q1, the fitted source luminosity is $\mathrm{Log(L_{\star}/L_{\odot})=5.2}$ . For Q9, the source luminosity is difficult to estimate because there is no J band flux measurement for this object. Based on the 2MASS detection limits, the models of Q9 would suggest that the stellar photosphere must be obscured in J band, implying $\mathrm{Log(L_{\star}/L_{\odot})=Log(L_{IR}/L_{\odot})=5.0}$ which is consistent with the non-detection of photospheric lines in the near-IR by Geballe et al. (2014). Although, if T$_{\star}$ is hotter than our models assume, the shift of the stellar spectrum to shorter wavelengths would imply a higher luminosity.

Determining grain sizes from fitting the SED alone are not well constrained because the shape of the SED is most strongly affected by total dust mass and the temperature distribution of the dust.  From a modeling standpoint, for a given grain size the distance from the star can adjusted to get the required temperature distribution to fit the SED.  Thus, measurements of the size of the emitting region add additional constraints. This effect can be easily seen from equilibrium heating of dust grains:

\begin{equation}
\mathrm{T_{eq}} = \Big( \frac{\mathrm{L}}{16 \pi \sigma d^2 } \frac{\mathrm{Q_{uv}}}{\mathrm{Q_{d}}} \Big)^{1/(4+\beta)}
\end{equation}

Where $\mathrm{T_{eq}}$ is the equilibrium temperature of the dust grain, $\mathrm{L}$ is the luminosity of the central source, $d$ is the distance from the central source, and $\sigma$ is the Stefan-Boltzmann constant. $\mathrm{Q_{uv}}$ is the grain-averaged absorption efficiency over the spectrum of the star, $\mathrm{Q_{d}}$ is the frequency-averaged dust emission, and $\beta$ is the assumed power law dependence of $\mathrm{Q_d}$ on temperature. For fixed equilibrium temperature and source luminosity which are determined by the SED fit, the distance from the central source is determined by the ratio of $\mathrm{Q_{uv}}$ and $\mathrm{Q_d}$. The dependence of grain size on these quantities imply that for fixed temperature, larger grains will be closer to the central source than smaller ones.

The FWHM of the QPMS as measured by SAM provide an upper limit on the physical size of the emitting region of the nebula, since none of the QPMs are spatially resolved. Computing FWHM for the DUSTY models, we find small (0.01 um) grains can be ruled out for Q1 and Q2 since they produce models with FWHM that are much larger than the measured FWHM for these objects. The observed FWHM for these objects agree fairly well with the model FWHM predicted by the medium (0.1 $\mu$m) grain size; however, we cannot rule out the presence of larger grains.

For Q9, the compact physical size implied by the SAM measurements are not replicated by any of the grain sizes considered. However, the small grain models (0.01 $\mu$m) provide the most satisfactory fit to the SED of Q9. Models using larger grains suffer from increasing optical depth at infrared wavelengths which adversely affects the fit for this object. Physically, this is due to the fact that larger grains exist at closer radii in the model, which for a constant dust mass (set by the overall emission) increases the optical depth of the model. Although the FWHM of Q9 determined by the SAM measurements is not well replicated by the DUSTY models, it should be noted that the spatial scales probed by the SAM measurements do not sample the large extended dust component around Q9 which is observed in the full-pupil mode discussed in the next section. This may explain the apparent inconsistency between the SED fit and the (SAM determined) FWHM of Q9.

\subsection{QPM Morphology}

\begin{figure}[ht]
\centering
\includegraphics[width=78mm,scale=1.0]{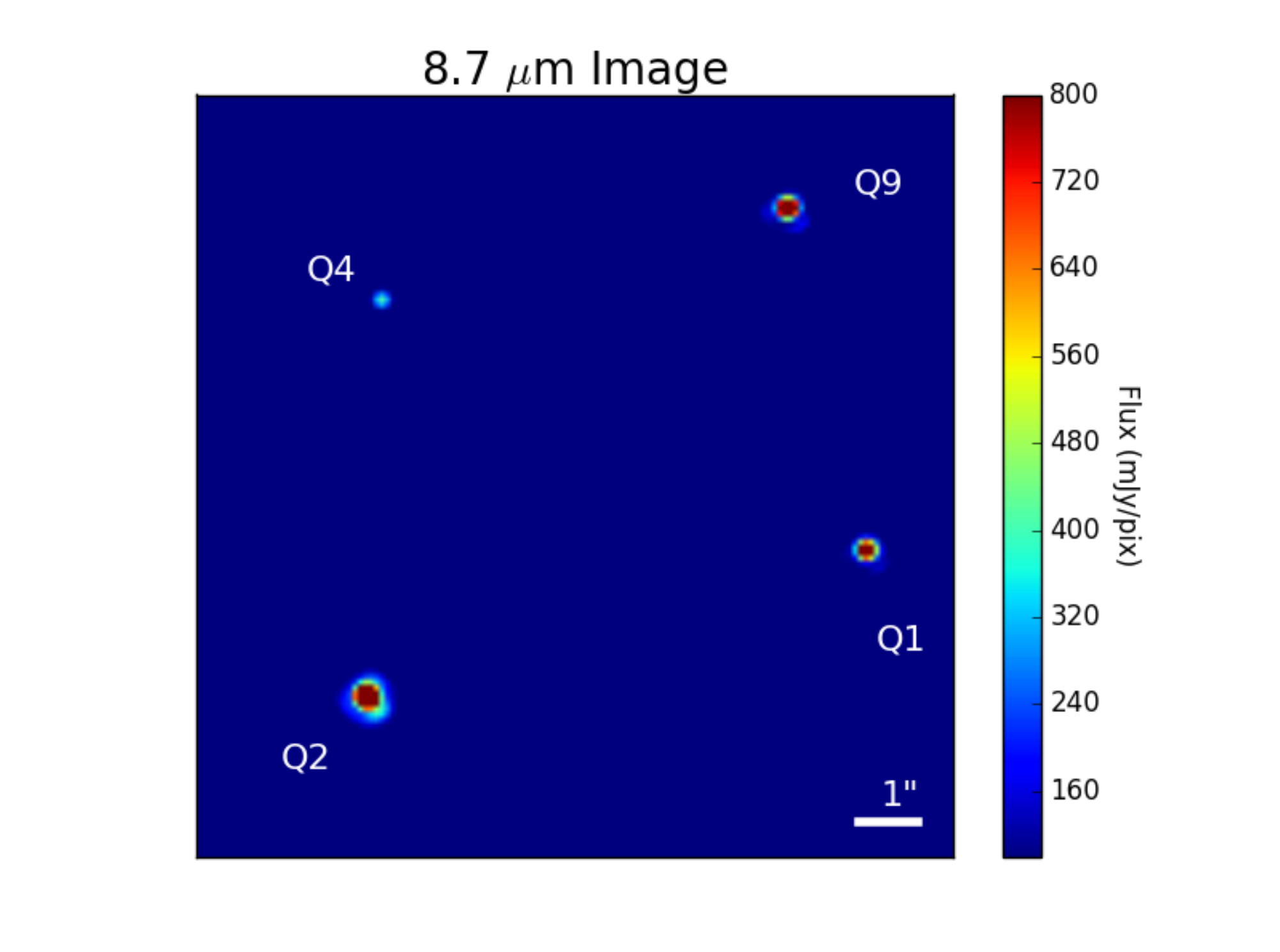}
\includegraphics[width=78mm,scale=1.0]{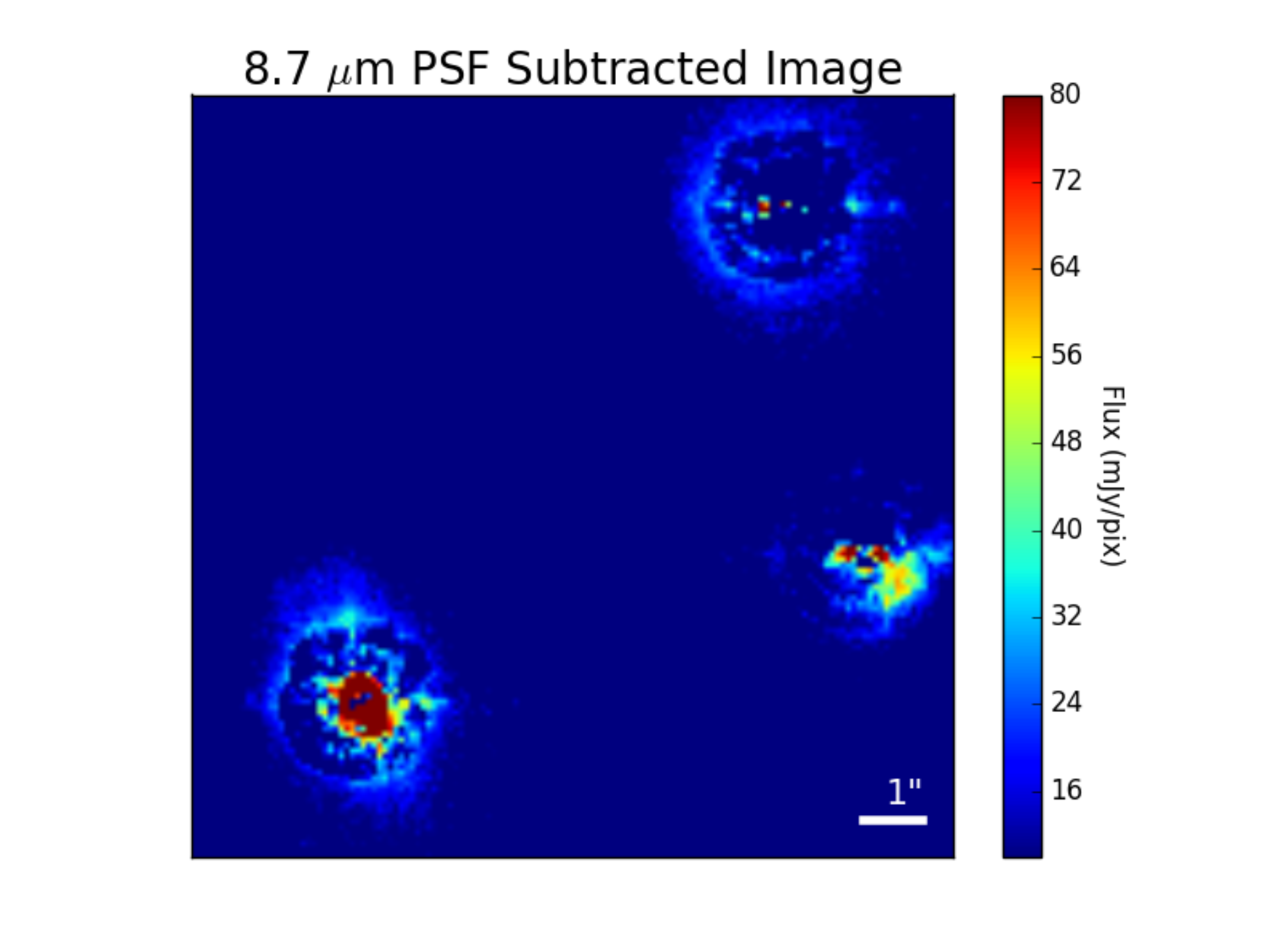}
\\
\includegraphics[width=78mm,scale=1.0]{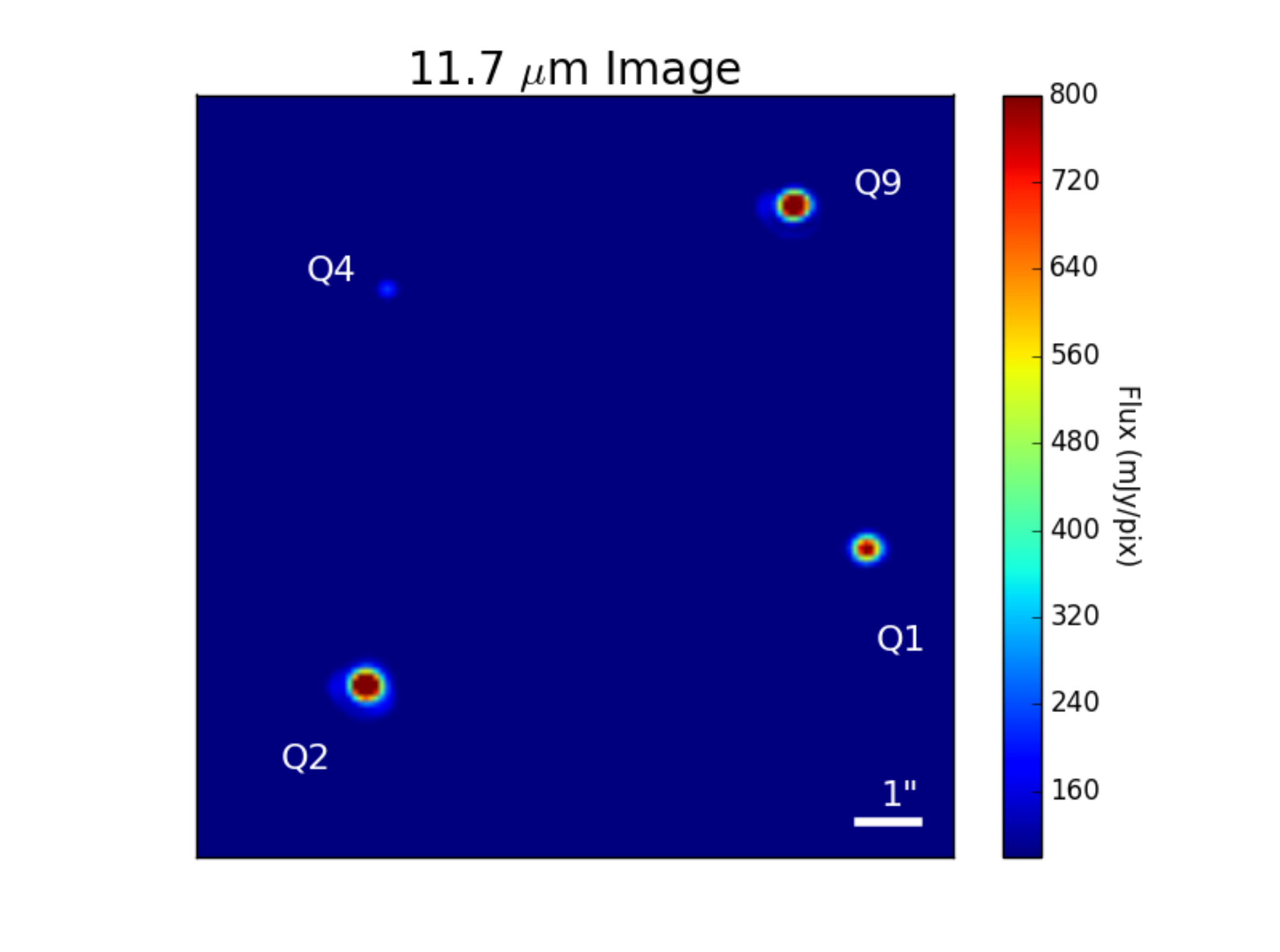}
\includegraphics[width=78mm,scale=1.0]{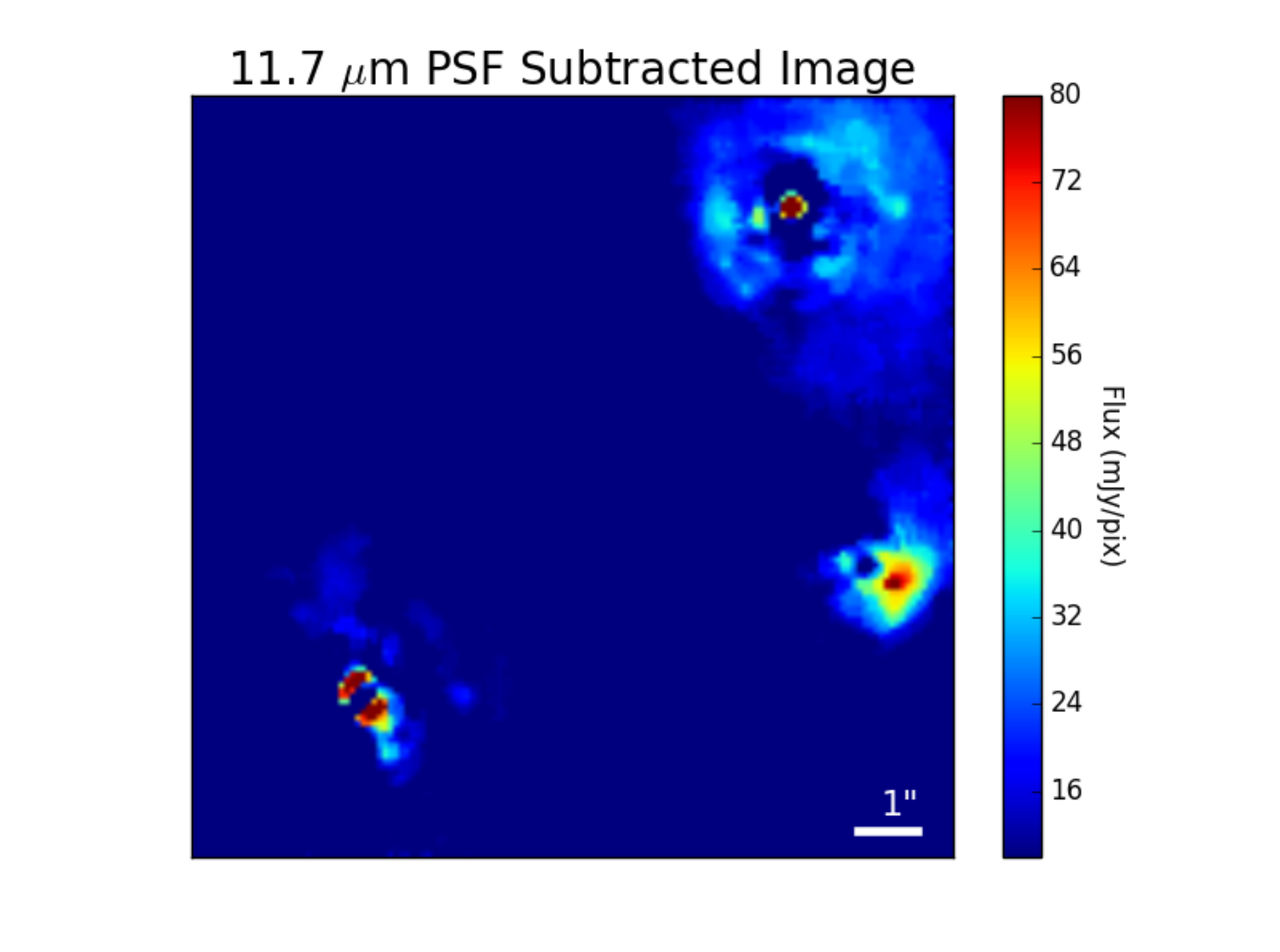}

\caption{Left: TReCS full-pupil data at 8.8 and 11.7 microns Right: PSF subtracted images using Q4 as the reference PSF. Q9 shows a significant amount of extended emission around the central source. Q1 shows some interesting emission to the southwest of the central source. Q2 shows north-south elongation of the central source which is consistent with the shape inferred from the SAM observations. In both subtracted images there is some low level diffuse extended emission around Q2 which may be artifacts left behind by the PSF subtraction process.}
\label{fig:fullpupil}
\end{figure}

The SAM images of Q2 (Figure \ref{fig:bsmem_im}) show spatially extended emission to the southwest of the main source. While this is at a level of only a few percent of the peak flux, the fact that it appears in our image reconstruction regardless of the prior model used, and that it occurs in the same location in both the 8.8 and 11.7 $\mu$m images, indicates that it is likely a real feature. Based on the size scales observed in the near-IR imaging of this system (Tuthill et al. 2006), this feature appears to be consistent with an outer portion of the pinwheel larger than the size scale of the innermost turn of the spiral structure ($\sim$100 mas). The extension in Q2 is also consistent with the North-South elongation that is observed in the full-pupil data of Q2 at 8.8 and 11.7 $\mu$m (Figure \ref{fig:fullpupil}). There is an additional over-resolved component in the SAM images for each of the QPMs. This could be due to structures in the nebulae that are larger than the mask allows to be sampled (the shortest baseline in the observations (1.75 m) corresponds to a maximum angular size of 1.4").

For Q1 and Q9, the full-pupil data are far more telling than the SAM data. The PSF subtracted images at 8.8 and 11.7 $\mu$m (Figure \ref{fig:fullpupil}) reveal asymmetric emission around Q1, which is strongest to the southwest. For Q9, there is a large $\sim 1-2"$ structure present in the 11.7 $\mu$m image which is only very faintly detected at 8.8 $\mu$m. From the imaging data alone, it is not clear that Q9 has a more extended pinwheel-like morphology or if it could be a shell of material. 

\begin{figure}[htp]
\centering
\begin{minipage}[l]{1.0\linewidth}
    \includegraphics[width=4 cm]{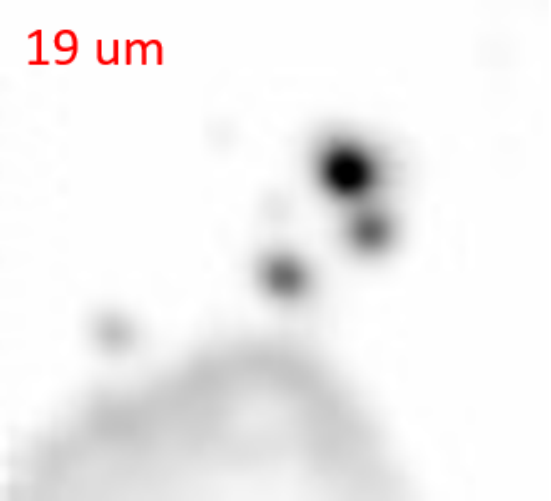}
    \includegraphics[width=4 cm]{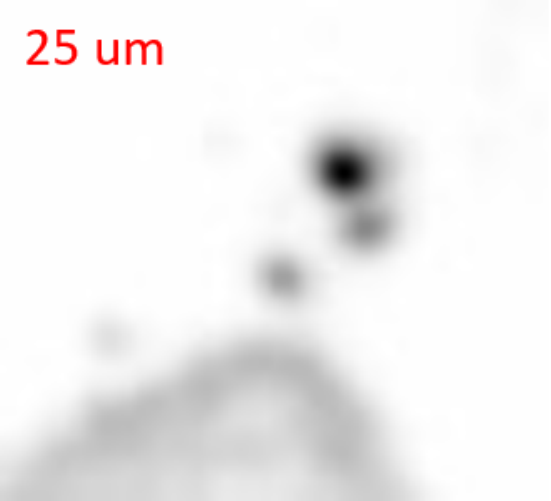}
    \includegraphics[width=4 cm]{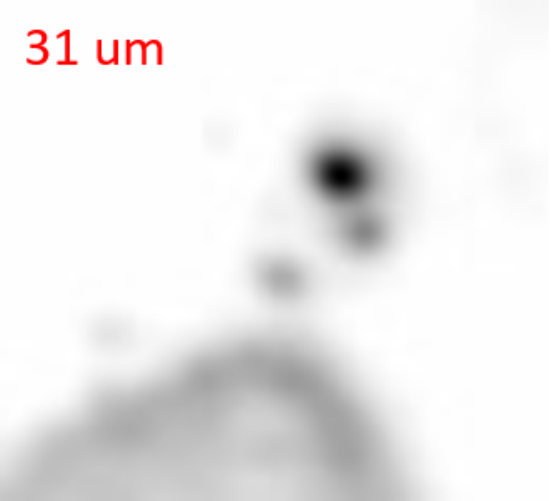}
    \includegraphics[width=4 cm]{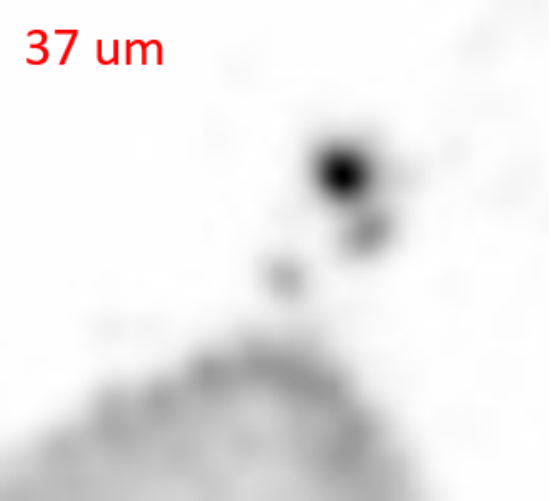}
\end{minipage}
\begin{minipage}[l]{1.0\linewidth}
    \includegraphics[width=4 cm]{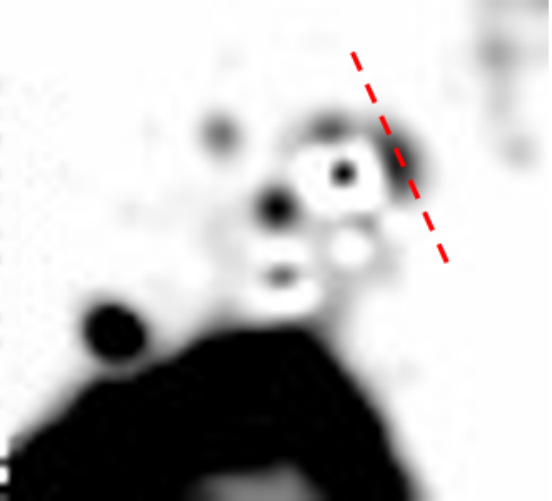}
    \includegraphics[width=4 cm]{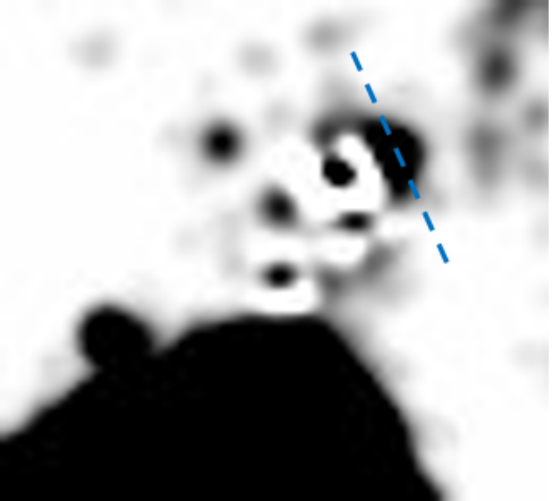}
    \includegraphics[width=4 cm]{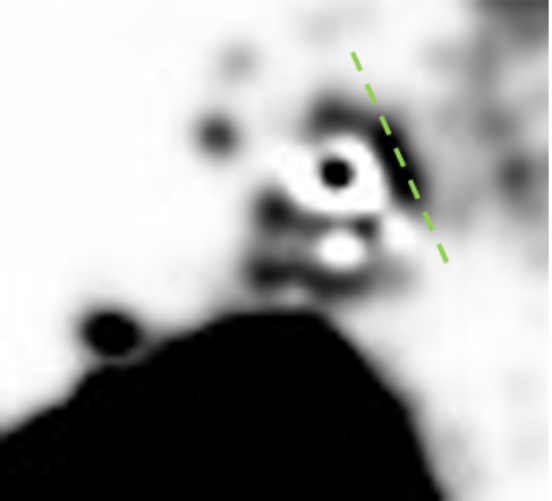}
    \includegraphics[width=4 cm]{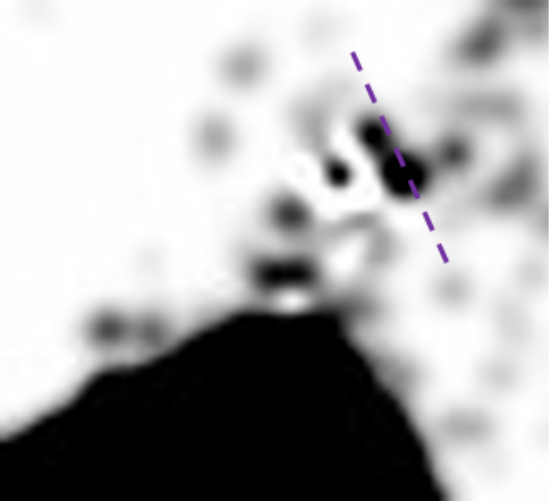}
\end{minipage}
\begin{minipage}[c]{1.0\linewidth}
    \vspace{5 mm}
    \centering
    \includegraphics[width=10 cm]{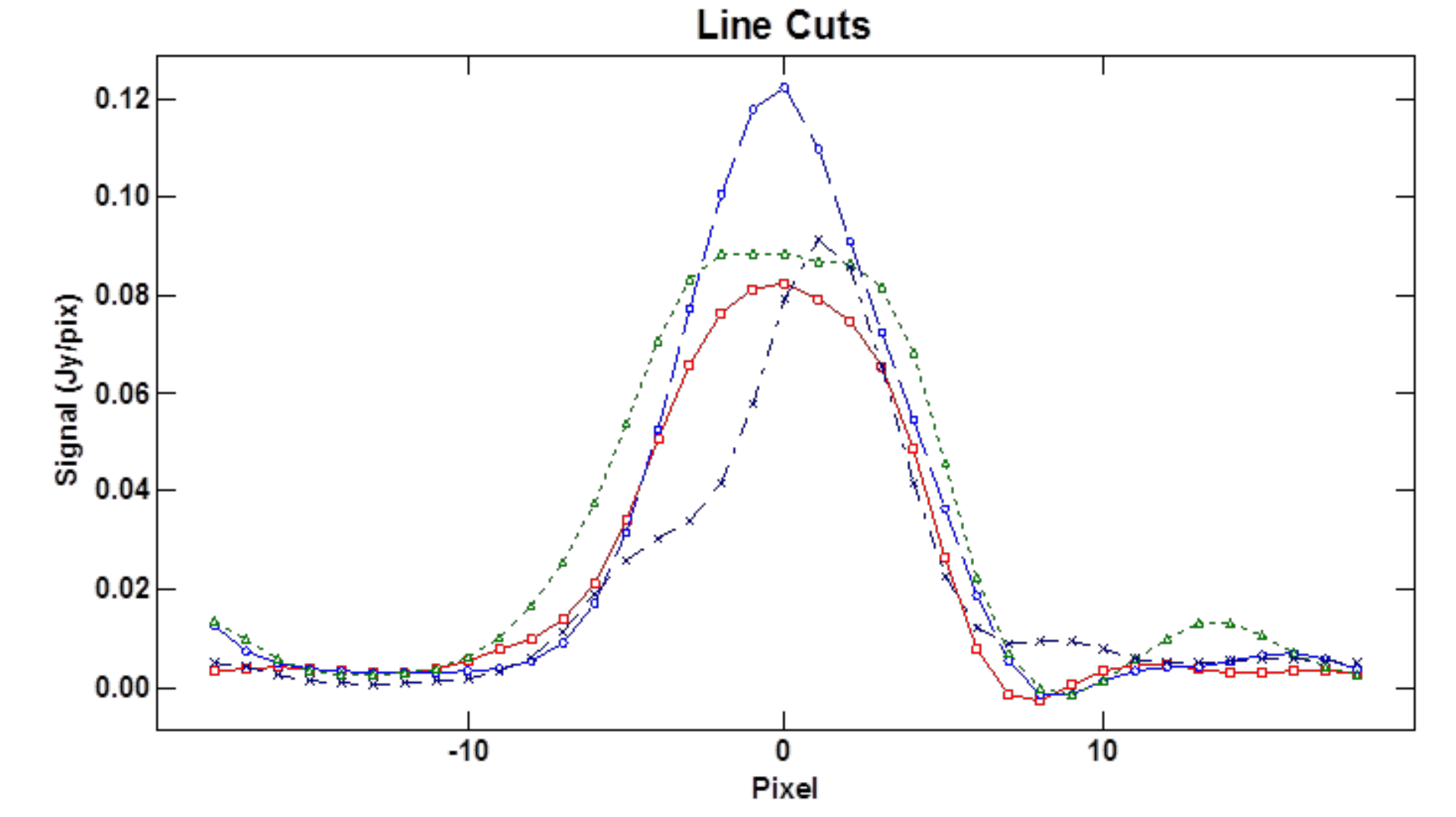}
\end{minipage}
  \caption{Top Row: FORCAST images at 19, 25, 31, and 37 $\mu$m. Second Row: FORCAST images with the components of the three brightest QPMs (Q9, Q1, and Q2) fitted by a 2D Gaussian and subtracted off from the image. A clear residual is left behind to the west of Q9. Bottom: Line cuts through the excess emission to the west of Q9; the line locations are shown in the second row. The intensity along the line is a 3 pixel sum tracing the main part of the emission. The colors of the lines correspond to: 19 (red), 25 (blue), 31 (green), and 37 (purple) $\mu$m. }
  %integrating under the curve we get: 3.9,4.5, 8.1, and 2.1 Jy for the wavelengths
\label{fig:forcastExcess}
\end{figure}

\begin{figure}[ht]
\centering
\includegraphics[width=80mm,scale=1.0]{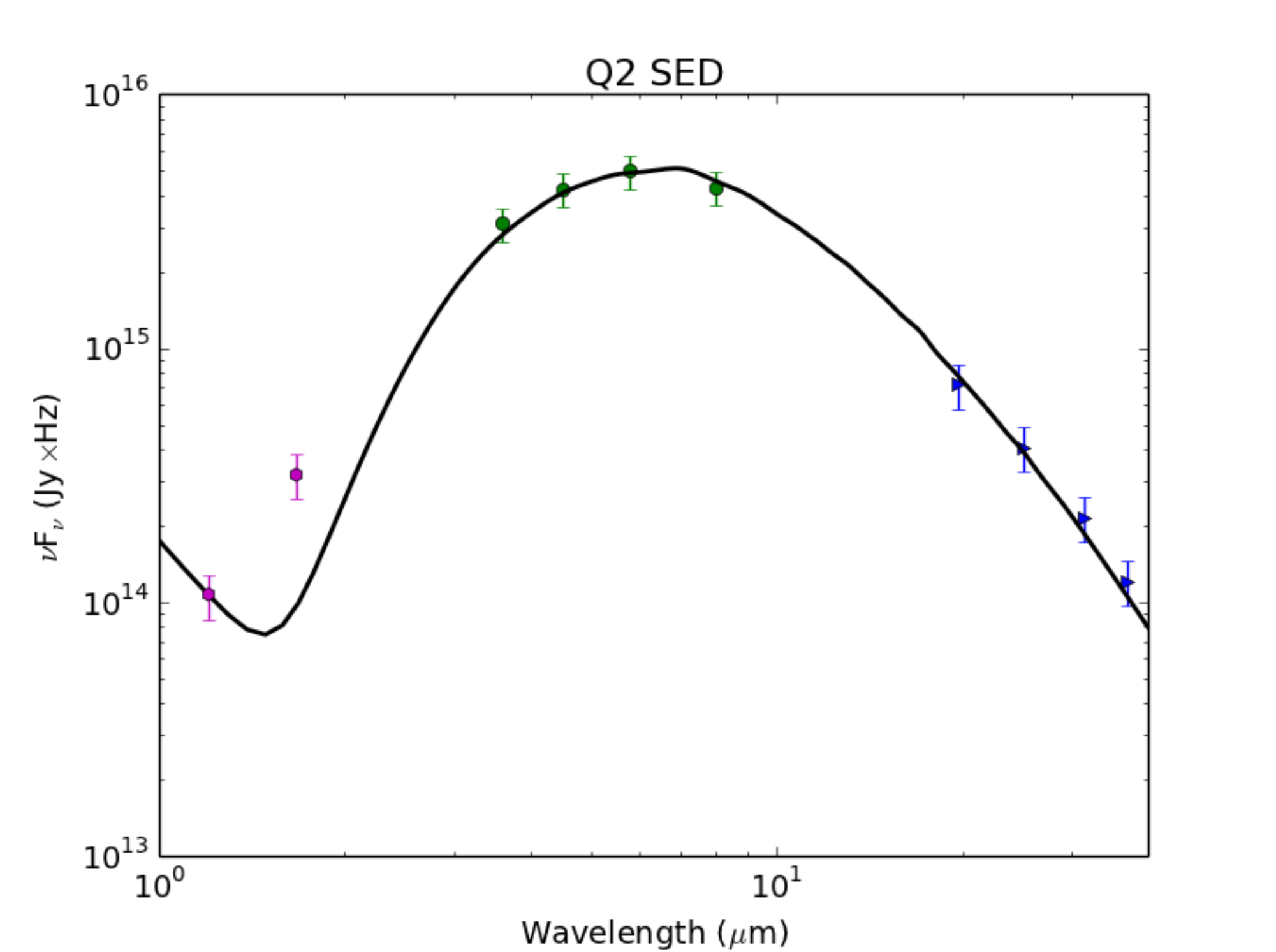}
\includegraphics[width=80mm,scale=1.0]{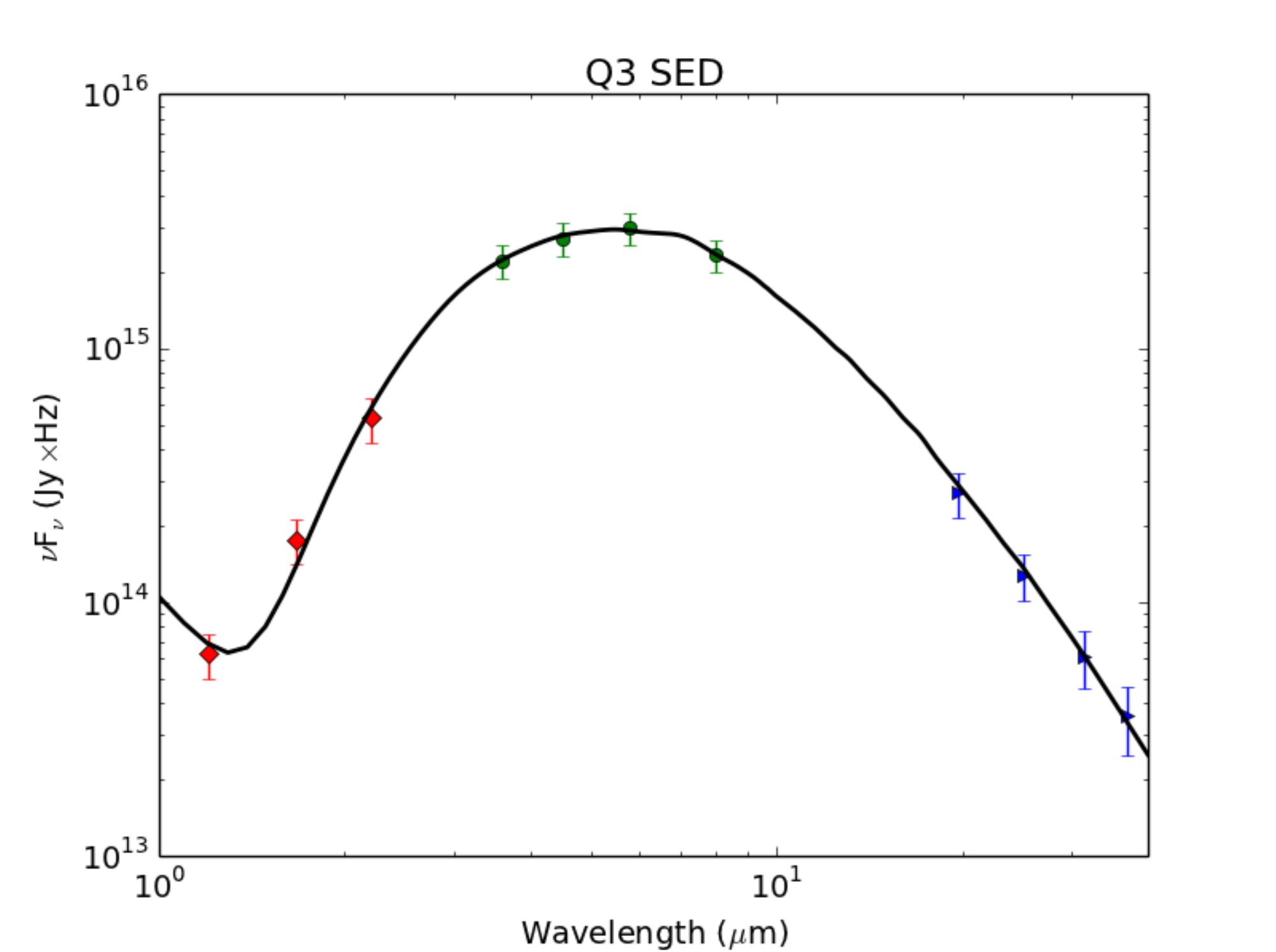}
\\
\includegraphics[width=80mm,scale=1.0]{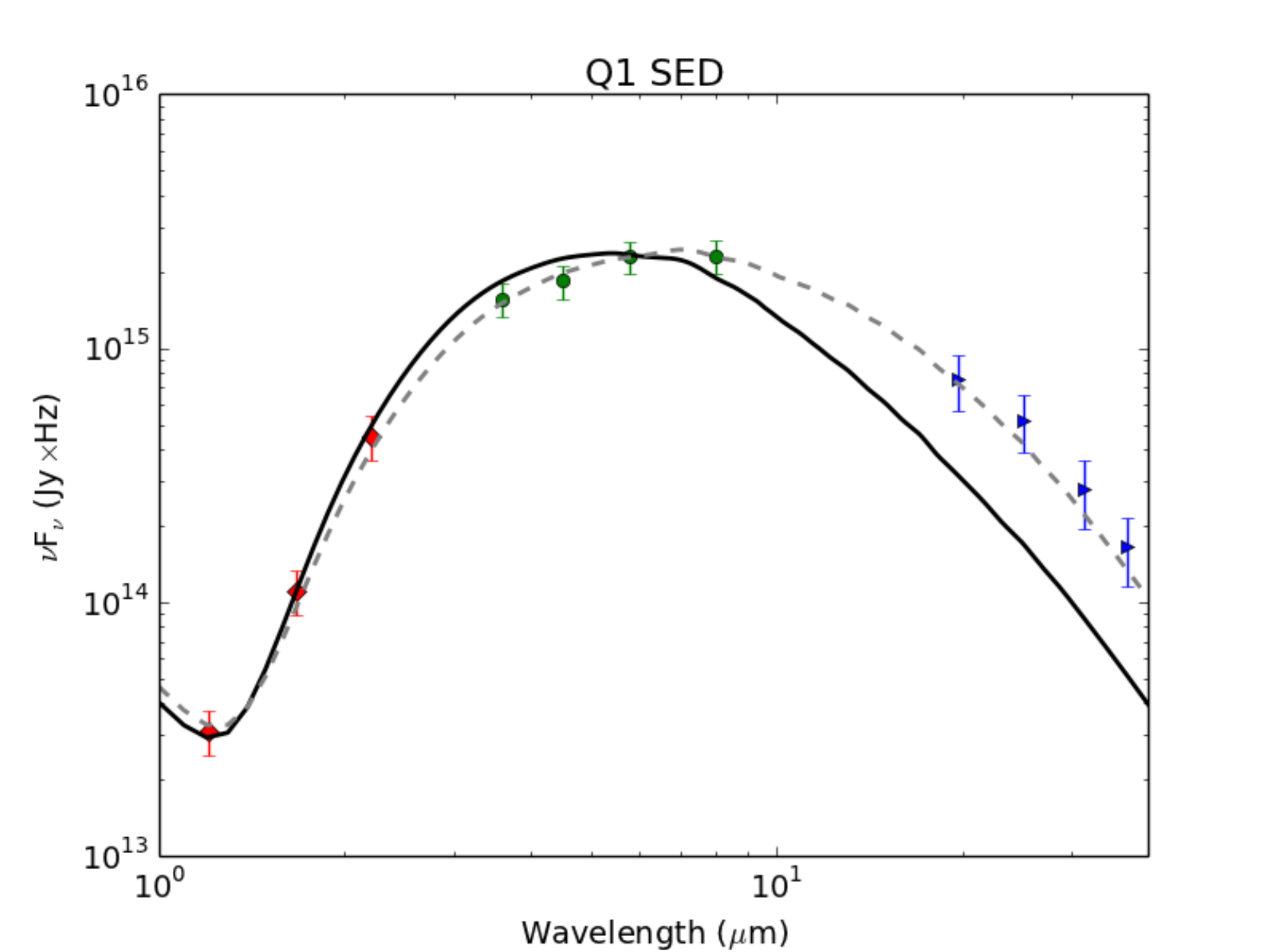}
\includegraphics[width=80mm,scale=1.0]{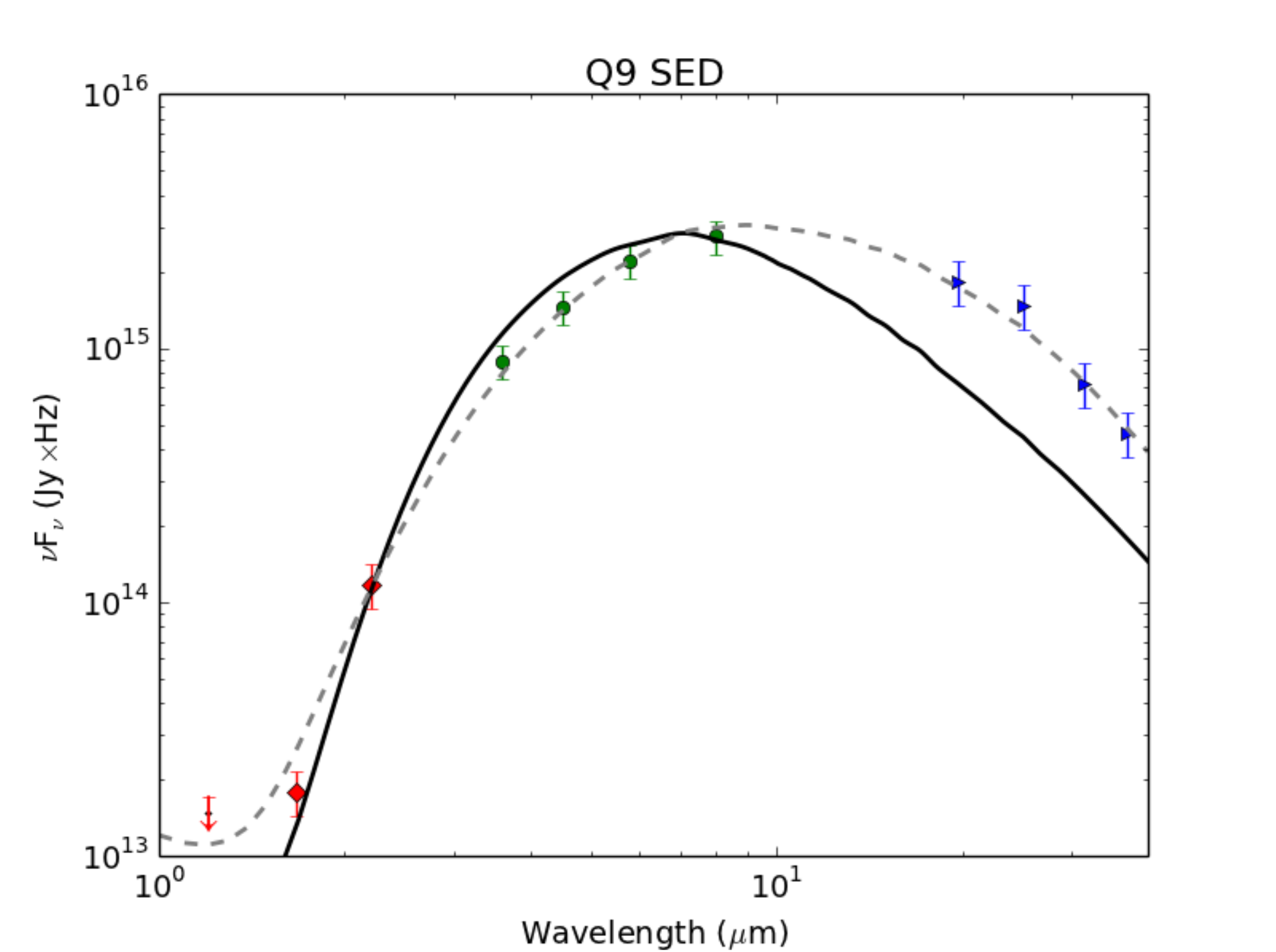}

\caption{SEDs and best-fit models for each QPM. The symbols are as follows: red diamonds are from 2MASS, green circles are from Spitzer/IRAC, blue triangles are from  SOFIA/FORCAST, and magenta hexagons are photometry from Matsunaga et al. (2009). The black solid curves our our best-fit $\mathrm{r^{-2}}$ density profiles. The dashed grey curves for Q1 and Q9 show our best-fit with an alternative density profile.  }
\label{fig:Qmodels}
\end{figure}

\begin{figure}[ht]
\centering
\includegraphics[width=100mm,scale=1.0]{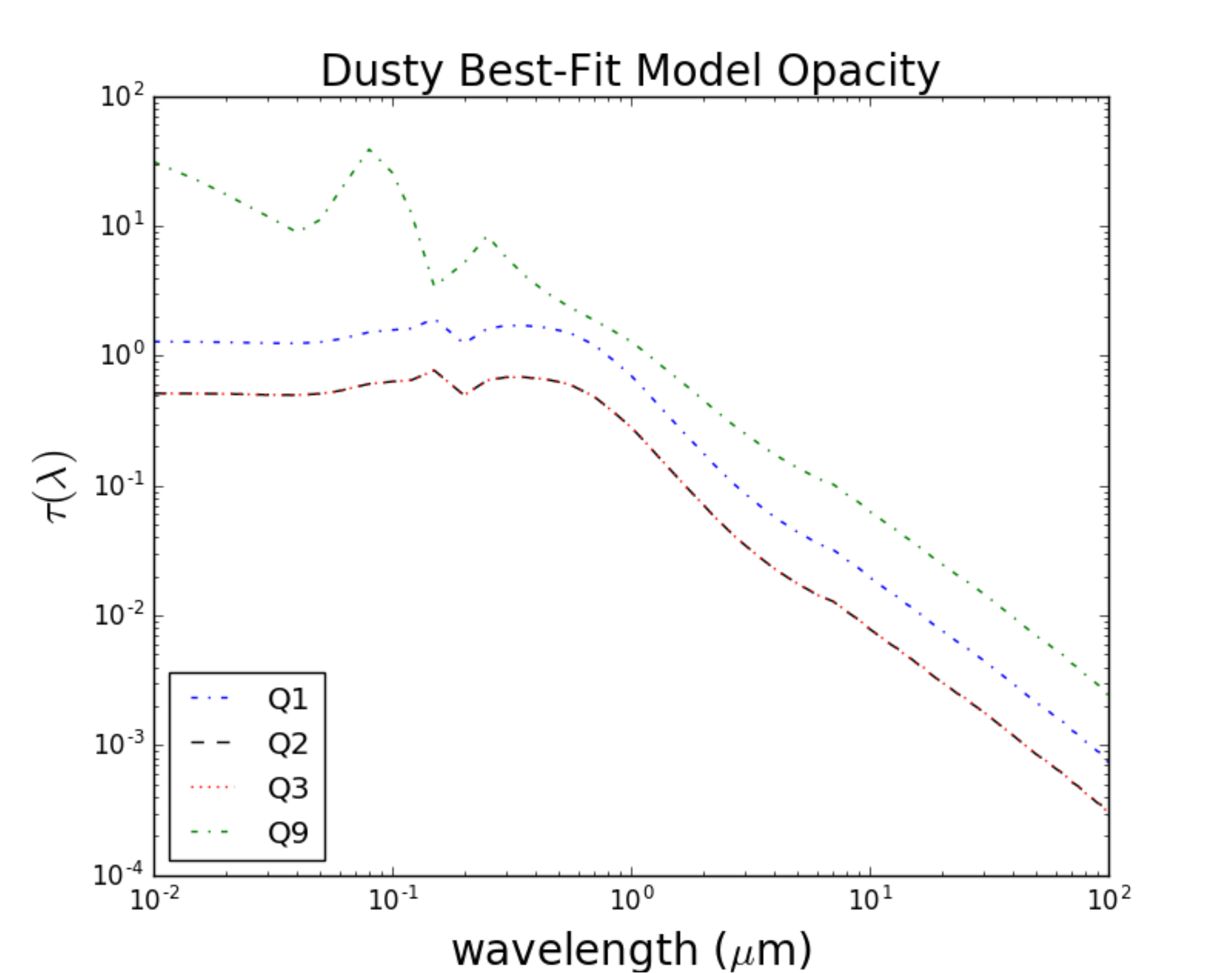}
\caption{Opacity of best-fit Dusty Models as a function of wavelength.}
\label{fig:tauPlot}
\end{figure}

In the FORCAST images, Q9 appears to be slightly spatially extended while the other QPMs appear point-like. The measured FWHM of Q9 at 19.7 and 37.1 $\mu\mathrm{m}$ are $4.6''\times 4.2''$ and $4.9''\times 4.4''$. Correcting for the beam size (3.4"), we calculate the deconvolved size of Q9 at 19.7 $\mu\mathrm{m}$: $3.2''\times 2.5''$ ($0.12\times 0.10$ pc) and at 37.1 $\mu\mathrm{m}$: $3.5''\times 2.8''$ ($0.14\times 0.11$ pc) with an estimated error of $0.3''$ (0.01 pc). The extended emission is similar to what is observed in the 11.7 $\mu\mathrm{m}$ image. In Figure \ref{fig:forcastExcess}, a Gaussian model was used to subtract off the fitted components from the brightest QPMs (Q9, Q1, and Q2) to show residual flux to the west of Q9 which is indicative its extended nature. The observed size of Q9 is not reproduced by the best-fit DUSTY models of this object: FWHM$_{19\mu m}$=0.86'' (0.033 pc) and FWHM$_{37\mu m}$=1.2'' (0.048 pc). Adjusting the various parameters considered in modeling this object does not improve the fit to the observed size scale, which indicates that something more is necessary to replicate this feature. For example, external heating by the Quintuplet cluster could potentially explain this discrepancy. Possible effects of the Quintuplet cluster on the QPMs are discussed in $\S$5.3.

\section{Discussion}

\subsection{IR Luminosity and Geometry}

The DUSTY models of the QPM SEDs provide slightly different results from those of Moneti et al. (2001) in terms of the infrared luminosity and the dust covering factor for the QPMs. The infrared luminosities, L$_{\mathrm{IR}}$, for the QPMs range from Log(L$_{\mathrm{IR}}$)=4.9-5.1 L$_{\odot}$.Taking the ratio of L$_{\mathrm{IR}}$ and L$_{\star}$, provides an estimate the geometric dust covering factors for the objects. For Q2 and Q3 the ratio of L$_{\mathrm{IR}}$ and L$_{\star}$ is 0.25 for both objects. The computed geometric covering factors are larger than typical disk-like geometries ($\sim$0.1). For Q1 and Q9 the ratio of L$_{\mathrm{IR}}$ and L$_{\star}$ are 0.5 and 1.0, respectively.

While the range of source luminosities for most of the QPMs is fairly typical for WC stars, the infrared luminosities are large in comparison. Williams et al. (1987) analyzed the population of known dusty WC stars ($\mathrm{N_{WC}}$=24) and found that only $\sim$20\% had dust covering factors greater than $\sim$0.1. It is interesting that the dust covering factors of all four objects are larger than this, which might suggest that the stellar luminosities are underestimated by our models. However, if the dust covering factors for the QPMs were $\sim$0.1 and we were to use the measured infrared luminosity to estimate the source luminosity, it would imply source luminosities that are factors of 4-6 larger than expected for late-type WC stars. Thus, the source luminosities fitted by the models, while poorly constrained, are more reasonable than our estimate based on a smaller dust covering factor. The extreme environment in which these objects reside could perhaps account for the large infrared luminosities. Possible environmental effects are investigated in $\S$5.3. In the case of Q9, the covering factor near unity and the preferred density profile are strongly suggestive that the emission may be from a shell of material rather than disk-like pinwheels such as those seen in Q2 and Q3. 

\subsection{Dust mass and Dust Production}

The observed dust mass in each of the QPMs ranges from $2.4\times10^{-5}-1.3\times10^{-3}$ M$_{\odot}$. The dust mass estimates are fairly robust (see Table \ref{tab:SpiralBestFit2}) since there is only a weak dependence on grain size. To estimate the total mass loss rates, we assume a typical WC wind velocity of 1500 km/s and a gas-to-dust ratio of 0.01. Given that these values could be reasonably off by factors of 2 and 3-5 respectively, the uncertainty in mass loss rates is about an order of magnitude.

If each of these systems fills the FORCAST beam ($\sim$3.4''), is located at the Galactic center ($\mathrm{d_{GC}}=8$ kpc), and has a wind velocity of 1500 km/s, this yields a timescale for the dust inventory of $\sim$40 years. Using this timescale and the assumed dust-to-gas ratio, the calculated mass loss rate of Q2 is $3.2\times10^{-4}$ M$_{\odot}/$yr. Since the QPM nebulae may not fully fill the beam, this estimate of the mass loss is likely a lower limit of the actual mass loss rate. For Q2, our estimate of the mass loss rate compares favorably with the mass loss rate determined by the radio emission of Q2 by Lang et al. (2005) who found $\dot{\mathrm{M}}=9.5\times10^{-5}$ M$_{\odot}$/yr. The two radio-derived and dust-derived mass loss rates become equal for a dust to gas ratio of 3\%. In a similar study of WCLds, Marchenko et al. (2002) found a dust to gas ratio of 6\% in WR 112. Q2 is the only QPM source for which a radio-derived mass loss rate exists. The remaining QPMs in this study have estimated dust-derived mass loss rates ranging from $6\times10^{-5}$-$3.2\times10^{-3}$ M$_{\odot}/$yr.

\subsection{Proximity to the Quintuplet Cluster}

Each of the QPMs appears to be $\sim$1 pc in projection from the center of the highly luminous Quintuplet cluster ($\mathrm{L_{QC}}=3\times10^{7}\mathrm{L}_{\odot}$; Figer et al. 1999). This cluster may provide an additional heating source for the dust or cluster winds may influence the dust distribution around the QPMs. Consider Q9, which is the lowest luminosity QPM ($\mathrm{L_{Q9}=10^5 L_{\odot}}$) and appears the most extended ($r_{\mathrm{Q9}}\sim0.1$ pc). At a distance of 0.08 pc from Q9, dust heating by the Quintuplet cluster would equal that of Q9 if it were located 1.4 pc away. Thus, if the Quintuplet cluster was close enough, it could cause the temperature profile of the outer regions to flatten out (in the 80-100 K range depending on grain size). We constructed a series of models of Q9 with a modified temperature profile to mimic this effect, however, fits to the SED were unsatisfactory, suggesting that the cluster heating is not likely significant. The other QPMs are less likely to be affected in this way since they are intrinsically more compact and luminous.

Winds from the Quintuplet cluster could directly shape the dust distributions around the QPM stars. The ram pressure balance between the winds in the Quintuplet and the winds in the central sources determine where the nebula might be disrupted by the nearby cluster:

\begin{equation}
\rho_{\mathrm{QC}}v_{\mathrm{QC}}^2=\rho_{\mathrm{QPM}}v_{\mathrm{QPM}}^2
\end{equation}

Where $\rho=\dot{\mathrm{M}}/4\pi r^2 v$ where $r$ is the distance from the central star in the QPM case and the distance from the cluster in the QC case and $\dot{\mathrm{M}}$ is the mass loss rate. It is assumed that $v_{\mathrm{QC}} \approx v_{\mathrm{QPM}}$, since the winds from the Quintuplet will be dominated by the WR stars present in the cluster. Solving this equation, yields $\mathrm{\dot{M}_{QC}}/d^2_{\mathrm{QC}}=\mathrm{\dot{M}_{QPM}}/r^2_{\mathrm{QPM}}$. The mass loss rate of an individual QPM compared to the cluster as a whole can be estimated using the measured mass loss rates of objects in the Quintuplet cluster (Rockefeller et al. 2005) which yields $\sim$0.1 of the total cluster mass loss for Q2. Thus, the stand-off distance from Q2 is 0.4 pc (using a cluster distance of 1.4 pc). The standoff radius for the other QPMs can be estimated by comparing the mass loss rates from Table \ref{tab:SpiralBestFit2}. Wind shaping is possibly significant for Q3, since the standoff radius of this object is the smallest (0.07 pc); however, the size scale is smaller FORCAST beam which is unresolved. Except for Q3, the standoff radius of the other QPMs are much larger than the observed extent of each object. This implies that wind shaping by the Quintuplet cluster is unlikely for most of the QPMs.

\subsection{The Unusual Nature of Q9}

Q9 appears to have physical properties that are very different from the other QPMs. In this section, we will discuss the various similarities and differences of Q9 with other dusty WC stars. Additionally, we will explore the possibility that Q9 is misidentified as a WC star.

There are several pieces of evidence that the extended dust emission associated with Q9 indeed originates from it. The observed dust in Q9 is purely carbonaceous in nature (Moneti et al. 2001; Chiar \& Tielens 2006) and thus unlike the general ISM (Draine 2003). Additionally, there is no reservoir of dust near the Quintuplet cluster as evidenced by the lack of submillimeter excess. To search for additional cool dust components that might be associated with Q9 we used the 70 $\mu$m  Herschel/PACS map of the region (Molinari et al. 2011). Due to the large beam size in the Herschel/PACS map ($\sim10''$), the individual QPMs are confused; however based on the fluxes and dust temperatures, Q9 is expected to dominate the emission at 70 $\mu$m. The measured flux from the QPMs at 70 $\mu$m (F$_{70}=17.1\pm1.2$ Jy) which agrees well with our model value of 16.1 Jy for Q9

If there is dust in present in the Quintuplet cluster, then it must be continuously replenished because of the powerful winds present in the region. Thus, it is highly unlikely that the dust emission associated with Q9 is due to a dense star-less clump since it would have been disrupted by the Quintuplet cluster on timescales much shorter than the lifetime of the cluster. Additionally, the dust near Q9 is warmer ($\sim 300$ K) than dust heated by the Quintuplet cluster alone ($\sim100$ K), implying that it must be heated by a central source.

If the reservoir of dust near Q9 was actually produced by Q9, the large physical size and cool dust temperatures observed in Q9 are similar to those of long-period colliding wind binary WC stars like WR 112 (P$\sim$25 years; Marchenko et al. 2002). Additionally, the luminosity of the central source is consistent with a late type WC star, as is the presence of amorphous carbon dust. However, to date, there have not been any spectral features identified in the near-IR which would provide strong evidence that Q9 is a late-type WC star. This is consistent with the DUSTY models of this object which indicate the star is too heavily dust obscured to detect the central source. While several properties of the dust emission of Q9 are consistent with other dusty WC stars, the mass of the reservoir and the implied mass loss rate are very large for typical WCLds. In a study of dusty nebulae around WC stars, Williams et al. (1987) found dust masses in these kinds of systems ranging from $1.2\times 10^{-9}$ to $1.9\times 10^{-4} \mathrm{M_{\odot}}$. In contrast, the observed dust mass present in Q9 ($\mathrm{M_d}=1.3^{+0.8}_{-0.4}\times 10^{-3}$ M$_{\odot}$) is much larger.

A possible cause of the large mass reservoir in Q9 is a lower than typical WC wind velocity which would allow for an increased timescale for the observable dust component to build up. This would also result in a smaller standoff radius between Q9 and the Quintuplet cluster which could make wind shaping of the nebula significant. For example, if the outflow velocity was 100 km/s, this would imply a standoff radius of 0.1 pc assuming the cluster located 1.4 pc away. This could potentially explain the observed asymmetry in the Q9 nebula, causing the near side of the nebula facing the Quintuplet cluster to be compressed while the far side would be allowed to expand freely. However, if Q9 is a colliding wind binary system as postulated for most dusty WC stars, the smaller wind velocities would likely be difficult to reconcile with the dust production mechanism in these types of systems. Alternatively, if Q9 is not a dusty WC star the assumed wind velocity and mass loss rate in this scenario are comparable to an eruption from a Luminous Blue Variable (LBV; Humphreys \& Davidson 1994) star. Although, outflows from LBVs are not typically carbon-rich as the dust in Q9 would suggest.

The SED fit for Q9 strongly prefers a much shallower density power law than one for a constant mass loss rate in time, suggesting that Q9 underwent a previous episode of enhanced mass loss in the recent past. Given our assumptions of typical WC star wind velocities, this would imply a typical timescale of the observable dust component of $\sim$40 years. Typical dusty WC star mass loss rates are an order of magnitude lower than what is inferred for Q9; however, given the relatively short timescale for dust sweep out in these kinds of systems, it is conceivable that that other dusty WC stars may also have brief periods of enhanced mass loss which are quickly swept away by their powerful winds on timescales of $\sim$ 100 years. However, with no other such events known, it is difficult to speculate on how common or uncommon these kinds of events may be. Ultimately, the differences between Q9 and other dusty WC stars make it an interesting and somewhat enigmatic object. Further study of Q9 is needed to constrain the nature of this object and how it relates to the rest of the dusty WC population. Additional high-resolution mid-IR mapping of the Q9 nebula could aid in further characterizing the nebula and determining the nature of this object.

\section{Conclusions}

In this paper, infrared observations taken with SOFIA/FORCAST and Gemini/TReCS were used to study the thermal dust emission from the QPMs. DUSTY models of the SEDs were used to determine properties of the dust reservoirs present in each object. Best-fit model parameters indicate that the near-IR pinwheel stars (Q2 and Q3) follow a $r^{-2}$ density profile, indicating a constant mass loss rate, while Q1 and Q9 show departures from a $r^{-2}$ density profile. High-resolution imaging shows large ($\sim$1'') extended structures associated with the latter two objects. Detailed modeling coupled with SAM observations favor the presence of larger dust grains than expected based on theoretical calculations in two of three objects where we have measurements. This is consistent with previous studies of dust grain sizes in other dusty WC stars. Based on the observed dust mass in these systems and the available radio mass loss rates, we derive a higher dust fraction than the ISM value in Q2 which is consistent with similar measurements of WR112. The massive dust reservoir in Q9 (M=$1.3\times10^{-3} \mathrm{M}_{\odot}$) is difficult to reconcile with the mass loss rates and wind speeds of typical dusty WC stars. The large dust reservoir and unusual density profile would suggest some kind of recent enhanced mass loss with $\dot{\mathrm{M}}\sim10^{-3}\mathrm{M}_{\odot}/$yr. Study of the literature on the WCLd population shows that periods of such enhanced mass loss in these kinds of objects are not typical, which possibly points to the highly unstable nature of massive stars near the end of their lives and the short timescales of the observable dust components in these types of systems. Further study of Q9 is needed to understand the nature of this very interesting object.

\emph{Acknowledgments} We would like to thank the rest of the FORCAST team, George Gull, Justin Schoenwald, Chuck Henderson, and Jason Wang, the USRA Science and Mission Ops teams, and the entire SOFIA staff. This work is based on observations made with the NASA/DLR Stratospheric Observatory for Infrared Astronomy (SOFIA). SOFIA science mission operations are conducted jointly by the Universities Space Research Association, Inc. (USRA), under NASA contract NAS2-97001, and the Deutsches SOFIA Institut (DSI) under DLR contract 50 OK 0901. Financial support for FORCAST was provided by NASA through award 8500-98-014 issued by USRA. This material is based upon work supported by the National Science Foundation Graduate Research Fellowship under Grant No. DGE-1144153. J. S.-B acknowledges support by grants AYA2012-38491-CO2-02 of the Spanish Ministry of Economy and Competitiveness (MINECO) cofounded with FEDER funds.

\clearpage

\vfill

\section*{Figures}

\section*{Tables}

\begin{table}[htp]
\centering
\footnotesize
%\caption{ Full-pupil and SAM Observations}
{\Large Table 1: Full-pupil and SAM Observations}\\
\vspace{2 mm}
\label{tab:obs}
\begin{tabular}{lllll}
\hline
Source & Filter & Data cubes & Frames & Data type\\
\hline
\hline
HD151680 & Si-2 8.74 $\mu$m & 2 & 102 & SAM calibrator\\
HD151680 & Si-5 11.66 $\mu$m & 2 & 102 & SAM calibrator\\
GCS3 & Si-2 8.74 $\mu$m & 2 & 102 & SAM \\
GCS3 & Si-2 8.74 $\mu$m & 2 & 102 & full-pupil \\
GCS3 & Si-3 9.69 $\mu$m & 1 & 102 & full-pupil\\
GCS3 & Si-4 10.38 $\mu$m & 2 & 102 & full-pupil \\
GCS3 & Si-5 11.66 $\mu$m & 2 & 102 & full-pupil \\
GCS3 & Si-5 11.66 $\mu$m & 2 & 102 & SAM \\
Eta Sgr & Si-2 8.75 $\mu$m & 2 & 102 & SAM calibrator \\
Eta Sgr & Si-5 11.66 $\mu$m & 2 & 102 & SAM calibrator \\
\hline
\end{tabular}
\end{table}

\begin{table}[htp]
\centering
\footnotesize
%\caption{ Best-fit Gaussian Model}
{\Large Table 2: Best-fit Faussian Model}\\
\vspace{2 mm}
\label{tab:best-fit}
\begin{tabular}{lcccccc}
\hline
 Parameters & Q2 8.8$\mu$m & Q2 11.7$\mu$m & Q9 8.8$\mu$m & Q9 11.7$\mu$m & Q1 8.8$\mu$m & Q1 11.7$\mu$m\\
\hline
\hline
FWHM [mas] & 105$\pm$4.45 & 128$\pm$7.29 & 58$\pm$6.39 & 88$\pm$6.02 & 73$\pm$3.48 & 126$\pm$5.01\\
$V^2_0$ (zero-spacing) & 0.81$\pm$0.01 & 0.69$\pm$0.02 & 0.89$\pm$0.01 & 0.71$\pm$0.01 & 0.9$\pm$0.01 & 0.78$\pm$0.01\\
\hline
\end{tabular}
\end{table}

\begin{table}[htp]
\footnotesize
\centering
%\caption{ BSMEM Parameters}
{\Large Table 3: BSMEM Parameters}\\
\vspace{2 mm}
\label{tab:bsmem_par}
\begin{tabular}{lcccccc}
\hline
 Parameters & Q2 8.8$\mu$m & Q2 11.7$\mu$m & Q9 8.8$\mu$m & Q9
                                                           11.7$\mu$m &
  Q1 8.8$\mu$m & Q1 11.7$\mu$m\\
\hline
\hline
Pixel scale& 25 & 35 & 25 & 35 & 25 & 35\\
Iterations & 20 & 30 & 17 &  27 & 17 & 20\\
$\chi^2$ & 2.04 & 2.17 & 1.07 & 3.82 & 1.02 & 1.21\\
\hline
\end{tabular}
\end{table}

\begin{table}[htp]
\footnotesize
\centering
%\caption{\centering Dereddened FORCAST Fluxes}
{\Large Table 4: Dereddened FORCAST Fluxes}\\
\vspace{2 mm}
\label{tab:forflux}
\begin{tabular}{ ccccc }
\hline
ID &  F$_{19.7 \mu m}$ (Jy) & F$_{25.2 \mu m}$ (Jy) & F$_{31.5 \mu m}$ (Jy) &  F$_{37.1 \mu m}$ (Jy)  \\ 
\hline
\hline
Q1 & 49.6$\pm$12.4 & 43.8$\pm$11. & 29.2$\pm$8.74 & 20.4$\pm$6.12 \\  
Q2 & 47.4$\pm$9.48 & 34.2$\pm$6.84 & 22.6$\pm$4.51 & 14.9$\pm$3. \\
Q3 & 17.7$\pm$3.54 & 10.7$\pm$2.14 & 6.42$\pm$1.93 & 4.4$\pm$1.3 \\ 
Q9 & 120.$\pm$24. & 124.$\pm$24.8  &  76.1$\pm$15.2  & 57.3$\pm$11.5 \\
\hline
\end{tabular}
\centering
%\caption*{{\footnotesize 3-$\sigma$ errors for the observations are 20\%. Q1 \& Q3 were accessed slightly higher errors due to their extraction. The following reddening factors were used for the data: A$_{19}$=0.89, A$_{25}$=0.63, A$_{31}$=0.40, and A$_{37}$=0.29}}
%\tablenotemark{ $3-\sigma$ errors for the observations are 20$\%$. Q1 $\&$ Q3 were accessed slightly higher errors due to their extraction. The following reddening factors were used for the data: A$_{19}$=0.89, A$_{25}$=0.63, A$_{31}$=0.40, and A$_{37}$=0.29}
\\{\footnotesize 3$\sigma$ errors for the observations are 20\%. Q1 \& Q3 were accessed slightly higher errors due to their extraction. The following reddening factors were used for the data: A$_{19}$=0.89, A$_{25}$=0.63, A$_{31}$=0.40, and A$_{37}$=0.29}
\end{table}

\begin{table}[ht]
\footnotesize
\centering
%\caption{\centering Best-Fit DUSTY Model Parameters}
{\Large Table 5: Best-Fit DUSTY Model Parameters}\\
\vspace{2 mm}
\label{tab:SpiralBestFit}
\begin{tabular}{cccccccc }
 %\multicolumn{9}{c}{\bf \large Table 3: Best Fit Spiral Model Parameters} \\
 \hline
ID & T$_{0}$ (K) &  r$_{0}$ (AU) & R & $\alpha$ (r$^{-\alpha}$) & $\tau_{\mathrm{J}}$ & a ($\mu$m) & $\chi^2$ \\ 
\hline
\hline
Q1 & 800$\pm$50 & 190$\pm40$     & 60$\pm20$  & 1.$\pm$0.5 & 0.5  & 0.1 & 1.2 \\
Q2 & 650$\pm$50 & 630$\pm140$   & 30$\pm$10  & 2 & 0.2 & 0.1 & 3.4 \\ 
Q3 & 750$\pm$50 & 330$\pm$60   & 30$\pm$15  & 2 & 0.2 & 0.1 & 1.2 \\
Q9 & 750$\pm$80 & 480$\pm$120   & 20$\pm$10   & -1.5$\pm$0.5 & 1.0  & 0.01 & 1.7 \\
\hline
\end{tabular}
\centering
\\{\footnotesize The best-fit parameters for DUSTY models of the QPMs. The parameter R is the non-dimensional radial size given in DUSTY which is a multiple of the inner boundary radius r$_0$. }
\end{table}

\begin{table}[ht]
\footnotesize
\centering
%\caption{\centering Additional Best-Fit Parameters}
{\Large Table 6: Additional Best-Fit Parameters}\\
\vspace{2 mm}
\label{tab:SpiralBestFit2}
\begin{tabular}{cccccc }
 \hline
ID & Log(L$_{\mathrm{IR}}$/L$_{\odot}$) & Log(L$_{\star}$/L$_{\odot}$) & L$_{\mathrm{IR}}$/L$_{\star}$ & M$_d$ ($\mathrm{M}_{\odot}$) & $\dot{\mathrm{M}}$ ($\mathrm{M}_{\odot}$/yr)\\ 
\hline
\hline
Q1 & 4.9 & $5.2$ & $\sim$0.5 &  3.8$^{+1.1}_{-1.5}\times 10^{-4}$ &  9.5$\times 10^{-4}$ \\
Q2 & 5.1 & $5.7$ & $\sim$0.25 &  1.3$^{+0.6}_{-0.5}\times 10^{-4}$ &  3.2$\times 10^{-4}$  \\ 
Q3 & 4.9 & $5.5$ & $\sim$0.25 &  2.4$^{+1.3}_{-0.8}\times 10^{-5}$ &  6.0$\times 10^{-5}$ \\
Q9 & 5.0 & $5.0$ & $\sim$1.0 &  1.3$^{+0.8}_{-0.4}\times 10^{-3}$ &  3.2$\times 10^{-3}$  \\
\hline
\end{tabular}
\centering
\end{table}

\begin{table}[ht]
\footnotesize
\centering
%\caption{\centering Grain Size \& Model FWHM}
{\Large Table 7: Grain Size \& Model FWHM}\\
\vspace{2 mm}
\label{tab:GrainSize}
\begin{tabular}{cccc}
 \hline
ID & a($\mu$m) & 8.8 $\mu$m FWHM (mas) & 11.7 $\mu$m FWHM (mas)\\ 
\hline
\hline
Q1 & 0.01 & 264$^{+92}_{-92}$ &  288$^{+100}_{-100}$ \\
   & 0.1 & 100$^{+35}_{-35}$ &  108$^{+38}_{-38}$ \\
   & 1.0 & 25.6$^{+9}_{-9}$ &  27.2$^{+10}_{-10}$ \\
\hline
Q2 & 0.01 & 202$^{+70}_{-70}$ & 211$^{+74}_{-74}$ \\ 
   & 0.1 & 171$^{+60}_{-60}$ & 176$^{+62}_{-62}$ \\ 
   & 1.0 & 48.1$^{+17}_{-17}$ & 46.9$^{+16}_{-16}$ \\ 
\hline
Q3 & 0.01 & 241$^{+84}_{-84}$ & 248$^{+87}_{-87}$ \\ 
   & 0.1 & 110$^{+39}_{-39}$ & 114$^{+40}_{-40}$ \\ 
   & 1.0 & 28.1$^{+10}_{-10}$ & 26.7$^{+9}_{-9}$ \\ 
\hline
Q9 & 0.01 & 510$^{+180}_{-180}$ & 623$^{+220}_{-220}$ \\
   & 0.1 & 570$^{+200}_{-200}$ & 570$^{+200}_{-200}$ \\
   & 1.0 & 245$^{+86}_{-86}$ & 268$^{+94}_{-94}$ \\
\hline
\hline
\end{tabular}
\centering
\end{table}

\end{document}